\documentclass[sigconf]{acmart}

\AtBeginDocument{}

\usepackage{soul}
\usepackage{graphicx}
\usepackage{xspace}
\usepackage{amsmath}
\usepackage{subcaption}
\usepackage{makecell}
\usepackage[T1]{fontenc}
\usepackage{enumitem}
\usepackage{algorithm}
\usepackage{algorithmicx}
\usepackage[noend]{algpseudocode}
\usepackage[scaled=0.9]{beramono}

\usepackage{amssymb}
\usepackage{setspace} 
\usepackage{multirow}
\usepackage{flushend}
\usepackage{algpseudocode}
\usepackage{caption}
\newcommand{\acro}{\textsc{Blest}\xspace}

\algrenewcommand\algorithmicrequire{\textbf{Input:}}
\algrenewcommand\algorithmicensure{\textbf{Output:}}

\begin{document}

\title{\acro: \underline{B}\underline{l}azingly \underline{E}fficient BF\underline{S} using \underline{T}ensor Cores}

\author{Deniz Elbek}
\authornote{Corresponding Author}
\affiliation{
  \institution{Sabanci University}
  \city{Istanbul}
  \country{Turkey}}
\email{deniz.elbek@sabanciuniv.edu}

\author{Kamer Kaya}
\affiliation{
  \institution{Sabanci University}
  \city{Istanbul}
  \country{Turkey}}
\email{kaya@sabanciuniv.edu}
\renewcommand{\shortauthors}{Elbek and Kaya}

\newcommand{\deniz}[1]{{\color{red} Deniz: #1}}
\newcommand{\kamer}[1]{{\color{blue} K: #1}}

\begin{abstract}
Breadth-First Search~(BFS) is a fundamental graph kernel that underpins a wide range of applications. While modern GPUs provide specialised Matrix-Multiply-Accumulate~(MMA) units, e.g., Tensor Cores~(TC), with extremely high throughput, they target dense operations, making it non-trivial to exploit them for irregular, unstructured graph computations. In particular, fully utilising them for a BFS requires an efficient mapping of the edge operations onto TCs while avoiding redundancy, load imbalance, and synchronisation. We present \textsc{\acro}, a TC-accelerated framework that reformulates the pull-based BFS pipeline around a bitmap-oriented structure and a carefully engineered execution layout. \textsc{\acro} introduces Binarised Virtual Slice Sets~(BVSS) to enforce warp-level load balancing and to eliminate frontier-oblivious work assignment. To improve both memory efficiency and update locality across diverse graphs, we apply two complementary graph reordering strategies: a compression-oriented ordering for social-like graphs and a bandwidth-reducing ordering for non-social graphs. At the compute level, we develop a batched SpMSpV multiplication pattern that uses the bitwise TC tiles to handle dot products without wasting output entries, thereby reducing the number of required MMA calls. Finally, \textsc{\acro} combines kernel fusion with a lazy vertex update scheme to reduce host-side synchronisation, mitigate atomic overheads, and improve cache locality. Experiments show that \textsc{\acro} delivers, on average, $3.58\times$, $4.64\times$ and $4.9\times$ speedup over BerryBees, Gunrock, and GSWITCH, respectively, across a broad set of real-world graphs.
\end{abstract}


\received{20 February 2007}
\received[revised]{12 March 2009}
\received[accepted]{5 June 2009}

\maketitle

\section{Introduction}

Breadth-First Search (BFS) is an important kernel used in a wide range of domains, including computer networks~\cite{computer_networks}, navigation systems~\cite{navigation_systems}, network analysis~\cite{diameter_computation}, computational biology~\cite{computational_biology}, and recommendation systems~\cite{recommendation_systems}---its performance has a direct impact on end-to-end application throughput. A single-source BFS is simply an iterative sparse matrix--sparse vector~(SpMSpV) multiplication over the Boolean semiring in a pull-based formulation. The pull view aligns well with GPUs since it avoids the irregular expansion patterns of push-based frontiers. However, modern GPUs have shifted peak throughput toward specialised dense-math units, e.g., Tensor Cores~(TC), which provide remarkable performance for matrix-matrix multiplication, but do not have direct support for irregular sparsity. Consequently, to benefit from TCs, BFS must be emulated by recasting it as an SpMM, while preserving the Boolean semantics and avoiding redundant work. In this work, we batch many dot-products into TC tiles, effectively implementing a small SpMM-like primitive while still computing SpMSpV semantics.

The state-of-the-art in TC-based graph traversal has shown that bitmap-centric representations are effective, even though the TCs have not been used with full efficiency. As we will show, several barriers remain.
First, graphs are highly irregular, so a fixed unit-of-work structure often leads to severe load imbalance.
Second, traditional slice-set processing is frontier-oblivious, causing warps to be scheduled for inactive work.
Third, the cost of vertex updates can dominate runtime due to scattered atomics and poor cache behaviour.
Finally, the compression effectiveness of bitmap slices depends strongly on the input vertex order, necessitating robust and graph-aware reordering strategies.

In this work, we present \textsc{\acro}, a blazingly efficient BFS method that exploits TCs and addresses these challenges holistically.
\textsc{\acro} combines data structure and algorithmic novelties to align the BFS pipeline on TCs and GPU memory hierarchies. In particular:
\begin{enumerate}[leftmargin=*]
    \item We propose a novel Binarised Virtual Slice Sets~(BVSS) data structure tailored to TC execution that helps to achieve perfect load balance across warps during a BFS.
    \item We propose two reordering strategies selectively used on different graphs: (i) to increase the BVSS's compression ratio, and (ii) to reduce the number of L1/L2 cache lines fetched from lower levels of the memory hierarchy.
    \item We develop a bitwise multiplication pattern that enables BFS to be executed {\em optimally} on TCs without redundant MMA operations and intra-warp communication.
    \item We propose a lazy vertex update scheme that reduces the cost of atomic operations and significantly improves cache locality.
    \item We introduce a kernel-fusion strategy that removes host-side synchronisation and avoids kernel launch overheads.
    \item On average, \acro is $3.63\times$ and $3.58\times$ faster than a state-of-the-art, Tensor-Core-based BFS framework, respectively, on the GAP benchmark and a custom benchmark; and is more than $4.6\times$ faster than state-of-the-art, non TC-based kernels, on average, across a broad set of real-world graphs.
\end{enumerate}

The article is organised as follows: In Section~\ref{sec:ds}, we introduce our data structure, BVSS, focusing on load balancing in Sec.~\ref{sec:load_balance} and vertex reorderings in Sec.~\ref{sec:ordering}. Section~\ref{sec:algo} presents \acro's compute mechanics and optimisations that refine the BFS pipeline. Sec.~\ref{sec:mult_pattern} details our multiplication pattern that is optimal for TCs. Sec.~\ref{sec:lazy_vertex} introduces our lazy vertex update scheme that reduces the cost of atomics and improves cache locality. Sec.~\ref{sec:kernel_fusing} describes our kernel-fusion strategy, which eliminates host-side synchronisation, and in Section~\ref{sec:experiments}, we compare \acro against SotA BFS implementations. Section~\ref{sec:related_work} provides a discussion on the related work, and Section~\ref{sec:conclusion} outlines future research directions and concludes the paper.

\section{Background and Notation}
\label{sec:background}
Let $G = (\mathcal{V}, \mathcal{E})$ be a simple directed graph with no loops, $n = |\mathcal{V}|$ vertices, and $m = |\mathcal{E}|$ edges. In a frontier-based parallel BFS, starting with the source vertex $s \in \mathcal{V}$ as the initial frontier, the $k$th iteration discovers the {\em next} frontier which is exactly \(k\) hops away from \(s\). This continues until every vertex reachable from \(s\) has been discovered. The kernel can return either a BFS tree, i.e., a $parent$ array, where \(\textit{parent}[u]\) for \(u \in \mathcal{V}\) stores the previous vertex on a shortest $s \leadsto u$ path, or a level array, where \(\textit{level}[u]\) shortest $s \leadsto u$ path length.

 A next frontier's vertices can be discovered in two directions. In the {\em top-down} approach, a vertex \(u \in \mathcal{V}\) in the current frontier {\em pushes} its state of being {\em visited} to its unvisited outgoing neighbours \(v \in \mathcal{V}\), e.g., $\textit{level}[u] + 1 \rightarrow \textit{level}[v]$. On the contrary, in the {\em bottom-up} approach, an unvisited vertex $v$ {\em pulls} this information from its incoming neighbours in the frontier $u$, e.g., $\textit{level}[v] \leftarrow \textit{level}[u] + 1$. Beamer et al.\ introduced {\em direction-optimised BFS}~\cite{direction_optimized_bfs}, in which, at each level, the information is transferred either in {\em pull} or {\em push} form depending on the number of (outgoing) edges of the current frontier and the number of (incoming) edges of the unvisited vertices. When the former is smaller, the direction is set to top-down, and otherwise, it is set to bottom-up. This \emph{direction-switching} mechanism can significantly reduce the execution time on graphs with dense frontiers, such as social networks. However, it also requires the incoming and outgoing views to be available in memory, which can be a problem on memory-restricted devices such as GPUs. 

For memory efficiency, \acro leverages an (only) pull-based approach by expressing the BFS computation as a sequence of sparse matrix–sparse vector (SpMSpV) multiplications over a semiring. Although a pull-only approach may scan more connectivity, \acro amortises it by the TC throughput and frontier-aware scheduling. Let \(\mathbf{A} \in \{0,1\}^{n \times n}\) be the sparse adjacency matrix of the transposed graph, and \(\mathbf{x} \in \{0,1\}^n\) be a (sparse) bit-vector representing the current frontier, i.e., \(x_u = 1\) if and only if $u \in \mathcal{V}$ is in the frontier. The SpMSpV operation $\mathbf{y} = A \otimes \mathbf{x}$,
(where \(\otimes\) denotes multiplication over the semiring) produces a bit-vector \(\mathbf{y} \in \{0,1\}^n\) which is the encoding of the next frontier. After each iteration, \(\mathbf{x}\) is set to \(\mathbf{y}\). 
 This stops when \(\mathbf{y}\) has no set bits, i.e., when the frontier is empty.

Let \(\mathbf{a}_u \in \{0,1\}^n\) be the row of \(\mathbf{A}\) corresponding to vertex \(u\), i.e., \(\mathbf{a}_u\) encodes the incoming neighbors of \(u\). The {\em pull} for \(u\) can be written as a dot-product $y_u = \bigoplus_{v=1}^{n} \left( a_{u, v} \otimes x_v \right)$ over the semiring
where \(\otimes\) and \(\oplus\) are the semiring ``multiplication'' and ``addition'' operators, respectively. In the Boolean (bitwise) semiring \acro works on, these scalar operations are instantiated as $\otimes \leftarrow \land, \oplus \leftarrow \lor$ so that the multiplication is replaced by bitwise {\sc{and}}, and the addition is replaced by bitwise {\sc{or}}. Hence, \(y_u = 1\) if and only if there exists at least one incoming neighbour \(v\) of \(u\) such that \(a_{u, v} = 1\) and \(x_v = 1\).

Representing BFS as iterative matrix-vector multiplication has been explored before, particularly in kernels performing multiple BFSs, e.g., computing the diameter~\cite{diameter_computation}, multi-source shortest paths~\cite{ibfs}, centrality~\cite{centrality, 6969541, kaya_centrality}, and influence maximization~\cite{influence_maximization}. For the single-source setting, the pull logic has been explored~\cite{berrybees}. In a multi-source setting, each source yields a frontier, and stacking these column-wise yields a frontier matrix. 
Hence, the matrix-vector multiplication generalises to a sparse matrix–matrix (SpMM) multiplication. However, the core computation remains structurally similar to the single-source case, the main focus of this work.

\subsection{GPU Tensor Core Architecture}
Modern GPUs have specialised {\em Matrix-Multiply-Accumulate}~(MMA) units, also known as \emph{Tensor Cores}~(TC), which are ASIC blocks optimised for high-throughput dense matrix–matrix multiplication. TCs are heavily utilised in workloads where compute-bound kernels dominate the overall execution time, such as deep neural-network training and inference.

The Nvidia GPU architecture is organised into multiple compute units called \emph{Streaming Multiprocessors} (SMs), each divided into \emph{SM subpartitions} (SMSPs). An SMSP includes a warp scheduler capable of issuing one instruction per cycle to a \emph{warp}, a team of 32 threads that execute the same instruction in the {\em Single-Instruction Multiple-Thread} (SIMT) model. In recent GPUs, each SM contains multiple TC pipelines--typically one per SMSP, depending on the GPU.

TCs support multiple execution modes, covering a variety of data types, precisions, and even structured sparsity configurations. Although TCs can, in principle, operate on sparse matrices, the supported sparsity patterns are highly structured—an assumption that is rarely satisfied by matrices arising from real-world graphs, where sparsity is typically unstructured and irregular. A handful of studies that use the TCs in dense mode on sparse data to accelerate SpMV~\cite{bitmap_spmv, dasp_spmv}, SpMM~\cite{dtc_spmm, torsten_spmm, acc_spmm, brp_spmm, cute_spmm, flashsparse_spmm, groot_spmm}, triangle counting~\cite{triangle_counting}, reduction~\cite{reduction}, stencil~\cite{stencil}, quantized SpMM~\cite{quantized}, fully-homomorphic encryption~\cite{neo}, epistasis detection~\cite{epiclear}, graph neural networks~\cite{tc_gnn}, deep learning~\cite{deepneural_spmm}, and, more recently, BFS~\cite{berrybees}.

Let \(\mathbf{A}_{m \times k}\) and \(\mathbf{B}_{k \times n}\) be two tiles of bit-matrices whose product yields a result tile \({\mathbf C}_{m \times n}\). TCs support several dense bit-matrix MMA modes, notably \textit{m8n8k128}, \textit{m16n8k128}, and \textit{m16n8k256}~\cite{ptx}. Throughout this article, we choose the smallest tile configuration, \textit{m8n8k128}, to have fine-grain control over the computation. The tiles \(\mathbf{A}\), \(\mathbf{B}\), and \(\mathbf{C}\) are distributed across the 32 threads of a warp. Assuming each word is 32 bits, the input tile for \(\mathbf{A}\) and \(\mathbf{B}\) consists of 64 words in total distributed across the warp, so that each thread holds one word from \(\mathbf{A}\) and one word from \(\mathbf{B}\) in its two allocated registers. The output tile \(\mathbf{C}\) also consists of 64 words, where each thread stores two of these output words in an additional two registers.

Conceptually, in the Boolean semiring formulation discussed earlier, each entry of \({\mathbf C}\) is a single bit indicating whether the corresponding row (vertex) has at least one ``active'' incoming connection. Thus, in principle, 64 bits would suffice to store the entire result tile. However, the TCs do not directly implement the Boolean semiring with $\otimes \leftarrow \land, \oplus \leftarrow \lor.$ Instead, their bitwise MMA operates over a different semiring instantiation, $\otimes \leftarrow \land, \oplus \leftarrow +,$ where \(+\) is integer addition. For a given output entry \(c_{u,v}\) (row \(u\) of \(\mathbf{A}\), column \(v\) of \(\mathbf{B}\)), the TC effectively computes $c_{u,v}
  =
  \bigoplus_{w=1}^{k} \left( a_{u,w} \otimes b_{w,v} \right)
  =
  \sum_{w=1}^{k} \bigl( a_{u,w} \land b_{w,v} \bigr),$
i.e., it accumulates the number of set bits produced by the bitwise-{\sc{and}} operations along the row–column dot product. This is known as {\em popcount} accumulation, and each \(c_{u,v}\) is stored in a full 32-bit word to hold the integer sum.

\section{\acro Data Structure}
\label{sec:ds}

To enable Boolean BFS semantics, we reinterpret each \(c_{u,v}\) as
\begin{equation}
y_{u}
  \;=\;
  \begin{cases}
    1, & \text{if } \exists v \mbox{ s.t. } c_{u,v} > 0,\\[2pt]
    0, & \text{otherwise},
  \end{cases}
  \label{eq:bit}
\end{equation}

\noindent i.e., we threshold the popcount at \(1\). In other words, the TC runs over an ({\sc{and}}, $+$) semiring, and we post-process its integer outputs to emulate the ({\sc{and}},{\sc{or}}) semiring needed for the frontier updates.

Following the state-of-the-art on sparse computations on TCs, we leverage a bitmap data structure, analogous to those of~\cite{berrybees,dasp_spmv, bitmap_spmv} proposed for BFS and SpMV. \acro, similar to {\em Binarized Row Slice}~(BRS)~\cite{berrybees}, partitions the columns of the adjacency matrix \({\mathbf A}\) into consecutive {\em intervals} of size $\sigma$ , where the $i$th interval, $0 \leq i < \lceil n / \sigma \rceil$, covers the column range $\left[\sigma i, \sigma (i + 1)\right)$. For a given interval, any row with an incoming neighbour in this column range is represented as a (row) \emph{slice}. 
The slices for the $i$th interval form a \emph{slice set}, stored as tuples \((u, \mathit{mask_{u, i}})\), where \(u\) is the pulling row and \(\mathit{mask_{u, i}}\) is a \(\sigma\)-bit, nonzero integer where its $j$th bit \(\mathit{mask_{u, i}}[j] = \mathbf{A}[u][\sigma i + j]\). To keep track of the number of slices in each set, BRS maintains an array \(\textit{sliceSetPtrs}\) of size \(\lceil n / \sigma \rceil + 1\). The entries of this array play the same role as \(\textit{rowPtrs}\) in an ordinary Compressed Sparse Row (CSR): for a slice set  \(s\), the quantity $\textit{sliceSetPtrs}[s + 1] - \textit{sliceSetPtrs}[s]$ equals the number of slices in $s$.

\begin{figure*}[!t]
     \centering
     \includegraphics[width=\textwidth]{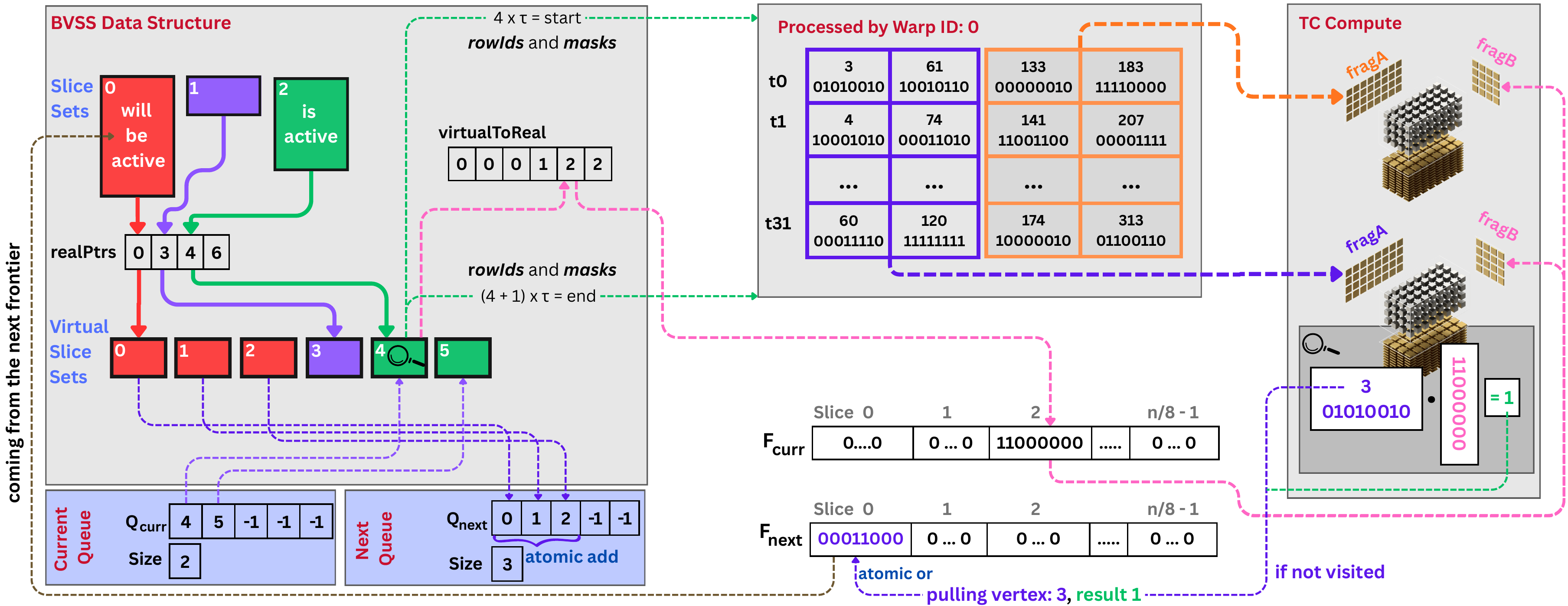}
    \caption{\small 
      BVSS data structure and the flow of data reads, pull operations, and frontier updates:
      Slice set~2 (in green) is active, with \textbf{F}$_{\text{curr}}$ bits {\tt 11000000}, i.e., the 1st and 2nd columns of slice set~2 are in the current frontier. Since \acro's queues operate over VSS indices, it retrieves the two VSSs corresponding to slice set~2, namely VSSs~4 and~5, directly from the current queue $\textbf{Q}_{\text{curr}}$ (bottom left). For simplicity, the figure focuses on VSS~4 whose slices ({\textit{rowIDs}} and {\textit{masks}}) are read from memory and assigned Warp~0. The masks and the $\sigma$ frontier bits of slice set~2 (accessed via \textit{virtualToReal} from VSS~4 and then from \textbf{F}$_{\text{curr}}$) are fed to the TC as {\textit{fragA}} and {\textit{fragB}}, respectively. Pulls on the Boolean semiring processed in two rounds, each processing half of the slices (colored purple and orange). For the slice with vertex/row~3 in the first half, the mask is {\tt 01010010}, indicating that vertex~3 is an outgoing neighbour of the 2nd, 4th, and 7th vertices/columns of slice set~2. Since the 2nd vertex is active in the frontier, the popcount is nonzero, and an update is required~(vertex~3 is unvisited). Since vertex~3 belongs to slice set~0 (covering vertices 0–7), its 4th bit is set in \textbf{F}$_{\text{next}}$. Finally, \acro locates all VSSs corresponding to slice set~0 via \textit{realPtrs} and inserts their IDs (0, 1, and~2) into $\textbf{Q}_{\text{next}}$.}
     \label{fig:data_structure}
     \Description{}
\end{figure*}

During multiplication, i.e., pull, each slice set $s$ is assigned to a single warp on the GPU. The bitwise (current) frontier vector \(\mathbf{x}\) is also partitioned into \(\sigma\)-bit words, so that \(\mathbf{x}[s]\) holds the frontier status of the $\sigma$ columns in \(s\). The warp iterates over all slices in $s$, and for each, it multiplies (on the semiring) the slice’s mask with \(\mathbf{x}[s]\). If this multiplication produces a nonzero (i.e., has at least one bit set) and the pulling vertex $u$ of the slice has not been visited before, the warp atomically adds $u$ to the next frontier.

Used by the state-of-the-art, this data structure and computation suffer from four major performance drawbacks:
\begin{enumerate}[leftmargin=*]
    \item \textbf{Load imbalance:} Real graphs are highly irregular; some vertices may have many outgoing edges, which yield crowded slice sets, while others remain relatively empty. Since each slice set is atomically mapped to a warp, this uneven slice distribution leads to load imbalance across the warps.
    \item \textbf{Frontier-oblivious set distribution:} Sparse frontiers are common in a BFS, whereas only a few frontiers are usually dense~(for road networks, frontiers are almost always sparse). The state-of-the-art assigns the slice sets to warps in a frontier-oblivious manner: each warp is assigned to a slice set $s$ regardless of whether \(\mathbf{x}[s]\) is zero or not. Even if a warp may detect a zero ($\sigma$-bit) word and exit, this frontier-oblivious assignment exacerbates load imbalance, since some warps may frequently exit early while others are repeatedly assigned to active ones.
    \item \textbf{Expensive vertex update cost:} When a pull from $u \in \mathcal{V}$ produces a nonzero popcount, and $u$ is unvisited, the thread performs an {\em atomic} update to insert $u$ into \textbf{F}$_{\text{next}}$. These updates are issued in the SIMT paradigm by the warp threads that satisfy the conditions. If the vertex IDs updated by the same warp are far apart, the resulting writes become highly uncoalesced, a fundamental factor that severely degrades GPU throughput.
    \item \textbf{Order-dependent compression:} The compression within the slices—storing 1 bit per edge—depends on the graph structure. Since connectivity is stored in \(\sigma\)-bit masks, the effective compression ratio can range from \(1/\sigma\) (nearly empty) up to \(\sigma/\sigma = 1\)~(fully utilised) bitmaps. To obtain high compression on a diverse set of graphs, an ordering scheme is indispensable.
\end{enumerate}

We will address the first three bottlenecks of the state-of-the-art in Section~\ref{sec:load_balance}, whereas the last one is addressed in Section~\ref{sec:ordering}.

\subsection{Binarised Virtual Slice Sets}
\label{sec:load_balance}
The state-of-the-art uses a \emph{slice set} as the smallest exploitable unit of parallelism, where each set contains all slices with intersecting incoming connectivity masks. To mitigate the impact of its flaws, \acro uses a novel data structure by which (i) load balance is enforced almost perfectly, and (ii) frontier-oblivious slice set distribution is avoided. One of the main ideas is partitioning the sparse matrix \(\mathbf{A}\) at the level of individual slices while still keeping the granularity of work assignment at the level of (active) slice sets. The proposed data structure, \emph{Binarised Virtual Slice Sets}~(BVSS), together with its usage, is illustrated in Figure~\ref{fig:data_structure}.

Let \(\tau\) be $\textit{WARP\_SIZE} \times {\textit{slicesPerThread}}$,
where \(\textit{WARP\_SIZE} = 32\) is the number of threads within a warp, and \(\textit{slicesPerThread} = 32 / \sigma\) is the number of slices, i.e., pull operations, each thread is assigned. Let \(\textit{numSlices}\) denote the total number of slices in the graph. \acro forms \emph{virtual slice sets}~(VSS) by enforcing two rules:
\begin{enumerate}[leftmargin=*]
  \item The number of slices in each VSS never exceeds \(\tau\).
  \item All slices in the same VSS must use the same $\sigma$-bit frontier word, i.e., they must be associated with the same set of columns. 
\end{enumerate}
These rules generate $numVSS$ virtual slice sets, which is at least \(\textit{numSlices} / \tau\). In \acro, the VSS is the smallest unit of work for concurrent computation. Moreover, on graphs with vertices having no outgoing edges, it is possible to have empty slice sets with no VSSs. In addition to the usual \(n\)-bit (current and next) \emph{frontier} arrays, and the size \(n\) \emph{level} array used to store the BFS tree, \acro leverages four static, read-only auxiliary arrays and two dynamic auxiliary arrays. The static arrays are explained as follows:
\begin{itemize}[leftmargin=*]
  \item \textbf{\textit{realPtrs}} (size = \(\lceil n / \sigma \rceil + 1\)) maps each real slice set to its corresponding range of VSSs. For a slice set \(s\), the quantity $\textit{realPtrs}[s + 1] - \textit{realPtrs}[s]$
  gives the number of VSSs associated with \(s\); these virtual slice sets are stored consecutively in memory.

  \item \textbf{\textit{virtualToReal}} (size = $numVSS$) maps each VSS back to its parent slice set, i.e., for a VSS \(\nu\), \(\textit{virtualToReal}[\nu]\) is the ID of the real slice set from which $\nu$ originates.

  \item \textbf{\textit{rowIds}} (size = \(numVSS \times \tau\)) stores the row IDs of all slices. Each row is stored as a 4-byte value.

  \item \textbf{\textit{masks}} $\left({\textnormal{size =}} \frac{numVSS \times \tau}{slicesPerThread}\right)$ stores the connectivity bitmaps of all (row) slices. Each element packs the $\sigma$-bit connectivity masks of \textit{slicesPerThread} slices as a compact, concatenated bitmap. 
\end{itemize}

\subsubsection{Processing VSSs in \acro}
The first dynamic array of the proposed BVSS is the queue \(\textbf{Q}_{\text{curr}}\) storing the VSS IDs to be processed at the current level. Similarly, the second array is the queue \(\textbf{Q}_{\text{next}}\) of VSSs for the next level. At each level, the VSSs in \(\textbf{Q}_{\text{curr}}\) are distributed to the warps via a round-robin assignment, i.e., the $i$th warp processes all the VSSs located at queue positions \(p\) such that $p \equiv i \bmod \textit{\#warps},$ where \(\textit{\#warps}\) is the total number of grid warps.

Suppose there exist two VSSs in \(\textbf{Q}_{\text{curr}}\) as illustrated in Fig.~\ref{fig:data_structure}. Let warp 0 be the warp assigned to the first VSS in the queue, which corresponds to VSS 4 coming from the slice set 0. Let $\sigma$ be 8, so that each thread is assigned \(\textit{slicesPerThread} = 32 / \sigma = 4\) slices. In this setting, a VSS with \(32 \times 4 = 128\) slices is processed in two rounds, each processing one half of the slices via Tensor Cores~(the reason is detailed in Sec.~\ref{sec:mult_pattern}). The first 2~(out of 4) slices assigned to each thread are handled in the first round. To read the appropriate $\sigma$-bit frontier entry of a VSS, \acro leverages \textit{virtualToReal} and maps the VSS ID to its slice set ID, used as an index to read from \textbf{F}$_{\text{curr}}$.\looseness=-1

After the multiplication/pull, each thread checks whether the vertices/rows associated with the two processed slices require an update; a row becomes active in the next level if and only if the corresponding output bit~(from Eq.~\eqref{eq:bit}) is \(1\) and it has not been visited before. For instance, the slice with row ID \(3\), which has not been visited before, yields the following two updates by thread \(0\):
\begin{enumerate}[leftmargin=*]
  \item \textbf{Update on \textbf{F}$_{\textnormal{next}}$:} The thread performs an {\tt atomicOR} on the 4th bit of the \textbf{F}$_{\text{next}}$\footnote{The update is on the 4th bit of the 1st word since row ID is 3 < 32.}. Since the hardware atomics operate at a 32-bit granularity, this locks the whole corresponding word in \textbf{F}$_{\text{next}}$.
  \item \textbf{Update on $\textbf{Q}_{\text{next}}$:} Knowing the ID of slice set 0 containing row 3, thread 0 identifies all the VSSs derived from slice set 0, which are at locations $[\textit{realPtrs}[0],\; \textit{realPtrs}[1])$. Since \acro uses VSSs as a unit of a task, thread 0 pushes the corresponding VSSs into $\textbf{Q}_{\text{next}}$ to process their slices in the next iteration. To reserve queue space, thread 0 atomically increments the queue size by \(\textit{realPtrs}[1] - \textit{realPtrs}[0]\) before enqueuing the new entries.
\end{enumerate}

Using  VSSs instead of slice sets balances the load of the warps; the number of TC pulls per warp is also well balanced. Thanks to the round-robin VSS distribution, the balance is close to optimal. An imbalance can only arise when the number of VSSs in $\textbf{Q}_{\text{curr}}$ is not divisible by $\#warps$. Even in this case, the difference in work between any two warps is bounded by two TC multiplications. 

The second problem, frontier-oblivious slice set distribution, is also resolved. \acro pushes the VSSs to $\textbf{Q}_{\text{next}}$ only when at least one of their columns is in the next frontier. Hence, warps are never assigned to VSSs with all-zero, $\sigma$-bit frontiers, i.e., except for the padded slices~(detailed in the following subsection), every unit of work corresponds to a genuinely active portion of the graph.

\begin{figure*}[htbp]
     \centering
     \includegraphics[width=\textwidth]{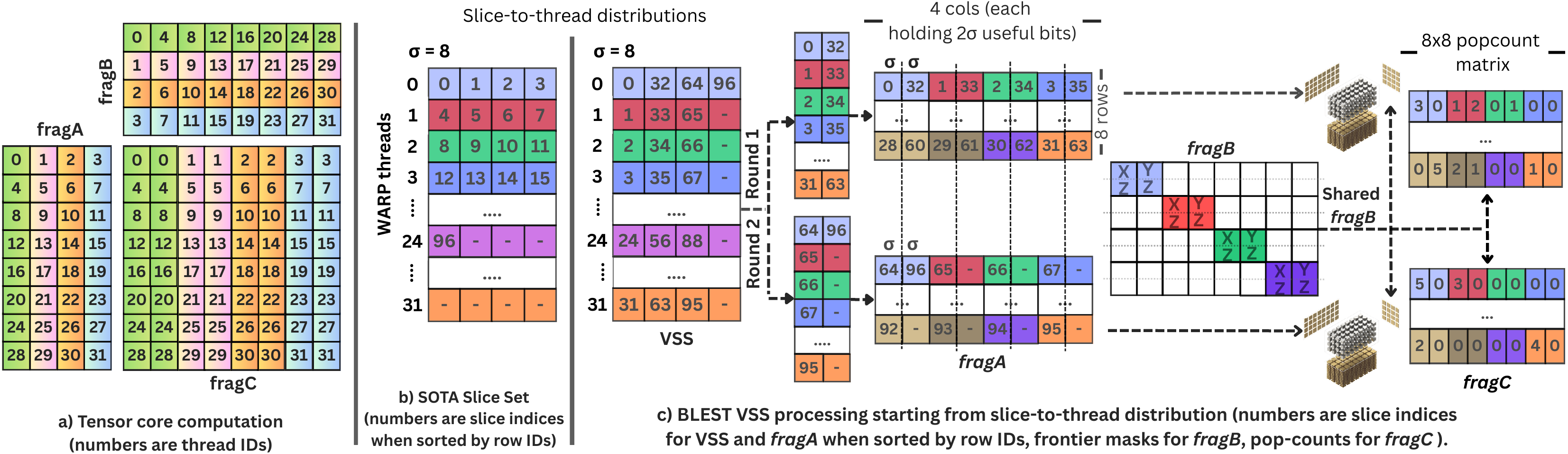}
     \caption{
     \small (a) The data layout for \textit{m8n8k128} on Tensor Cores. For the inputs {\textit{fragA}}, {\textit{fragB}} and the output {\textit{fragC}}, a box corresponds to a 32-bit word, and the number inside each box is the thread ID storing the corresponding word in its register. The figure assumes $\sigma=8$. (b) The slice-to-thread distribution of the SotA, which processes the current set's slices in four rounds, each performing four \textit{m8n8k128}s, i.e., 16 multiplications in total. (c) The slice-to-thread distribution of \acro reduces load imbalance and \textbf{F}$_{\text{next}}$ update divergence. The pull operations for the VSS at hand are handled in two rounds, each performing one \textit{m8n8k128}. For each stage, a 32-bit \textit{fragA} word contains two 8-bit masks coming from the same thread in its first 16 bits, followed by 16 zero bits~(omitted in the figure). Following this pattern, the 32-bit words of \textit{fragB} are all zeros except the ones corresponding to the threads $0, 4, 9, 13, 18, 22, 27$, and $31$. Let $\alpha$ be the 8 frontier bits for the current slice set. The 32-bit words have the form $XZ$ or $YZ$~(aligned vertically in the figure), where $X$, $Y$, $Z$ are 16-bit sequences; $X$ is $\alpha$ followed by eight 0s, $Y$ is eight 0s followed by $\alpha$, and $Z$ is all 0. Hence, when the 16 bits from \textit{fragA} are multiplied by $XZ$ or $YZ$, the first and the second 8 bits, respectively, are selected. The popcounts are obtained as an $8 \times 8$ matrix distributed to the warp as shown in part (a).}
     \label{fig:computation_structure}
     \Description{}
\end{figure*}

\subsubsection{Re-organising slices for a better performance}
\label{sec:reorganize}

Coalescing memory reads/writes within a warp boosts performance. In \acro, each thread \(t\) reads \textit{slicesPerThread} masks in a single 32-bit load and \textit{slicesPerThread} row IDs in a single 128-bit vectorised load, where the corresponding pulls are assigned to \(t\). With a similar mask-packing strategy, previous work~\cite{berrybees} keeps the slices sorted with respect to row IDs in the \textbf{\textit{rowIds}} and \textbf{\textit{masks}} arrays. This yields the slice-to-thread distribution matrix shown in Fig.~\ref{fig:computation_structure}(b), where \(\sigma = 8\), \(\textit{slicesPerThread} = 4\), and the numbers inside the boxes are the indices of the slices in their sorted order with respect to row IDs. The resulting \(32 \times 4\) distribution matrix is stored in row-major order so that each thread’s four slices (columns 0–3) are laid out consecutively in memory. However, due to SIMT execution, the intra-warp processing order is column-major: the kernel first processes column 0 for all threads and then proceeds up to column 3. Hence, all updates for a given column are issued together.

This slice organisation in~\cite{berrybees} leads to two problems. First, the updates are scattered and uncoalesced, yielding a large \emph{update variance}, as will be measured and detailed in Section~\ref{sec:ordering_cache_line}. Second, when the number of slices (97 in the figure) is not divisible by \(\tau\) (128 in the figure), the threads at the end of the warp may not be assigned any pull/update operations. To avoid these problems, \acro orders the slices in a VSS \(\nu\) using a different strategy. Let \(\textbf{\textit{masks}}[(\tau \times \nu) + i][j]\) denote the \(j\)-th \(\sigma\)-bit mask in the \(i\)-th 32-bit entry for \(\nu\), where each entry packs \textit{slicesPerThread} masks. Let \(k\) be the index of a mask in \(\nu\) when the slices are sorted with respect to their row IDs. Then each \(\sigma\)-bit sequence is placed as $\textbf{\textit{masks}}[(\tau \times \nu) + (k \bmod 32)][\lfloor k / 32 \rfloor] \leftarrow \textit{mask}.$ This generates the logical \(32 \times 4\) slice-to-thread distribution in Fig.~\ref{fig:computation_structure}(c, left).

When the number of slices in a VSS \(\nu\) is less than \(\tau\), the empty \(\sigma\)-bit locations in the 32-bit entries of \textbf{\textit{masks}} are padded with all-zero \(\sigma\)-bit masks. A similar padding is also applied to \textbf{\textit{rowIds}} with 32-bit random, dummy row IDs. Hence, in the proposed BVSS data structure, each VSS \(\nu\) is padded so that its number of slices is a multiple of \(\tau\). Note that this only happens when \(\nu\) is the last VSS of a slice set, as for VSSs 2, 3, and 5 in Fig.~\ref{fig:data_structure}. Otherwise, as for VSSs 0, 1, and 4 in Fig.~\ref{fig:data_structure}, \(\nu\) already contains \(\tau\) slices.

Since the processing occurs in column-major order and we sort the slices by row IDs in the same major order, we maximise coalescing within a warp during the processing of a VSS. Moreover, due to the padding strategy, where both useful and padded slices are distributed in round-robin fashion across the threads of a warp, we leave no threads without work unless the number of slices in the virtual slice set is smaller than \(\textit{WARP\_SIZE} = 32\).

\newcommand{\ordhead}[2]{%
  \begin{tabular}{@{}c@{}}
    \includegraphics[width=0.9cm]{#1}\\[0.2ex]
    \textbf{#2}
  \end{tabular}
}

\subsection{One Ordering Decision to Pull them All}
\label{sec:ordering}

\begin{table*}[t]
  \centering
  \scriptsize
  \setlength{\tabcolsep}{3pt}

\begin{subtable}[t]{0.5\linewidth}
  \centering
  \scalebox{1.3}{
  \begin{tabular}{lccccc}
    \textbf{} &
    \ordhead{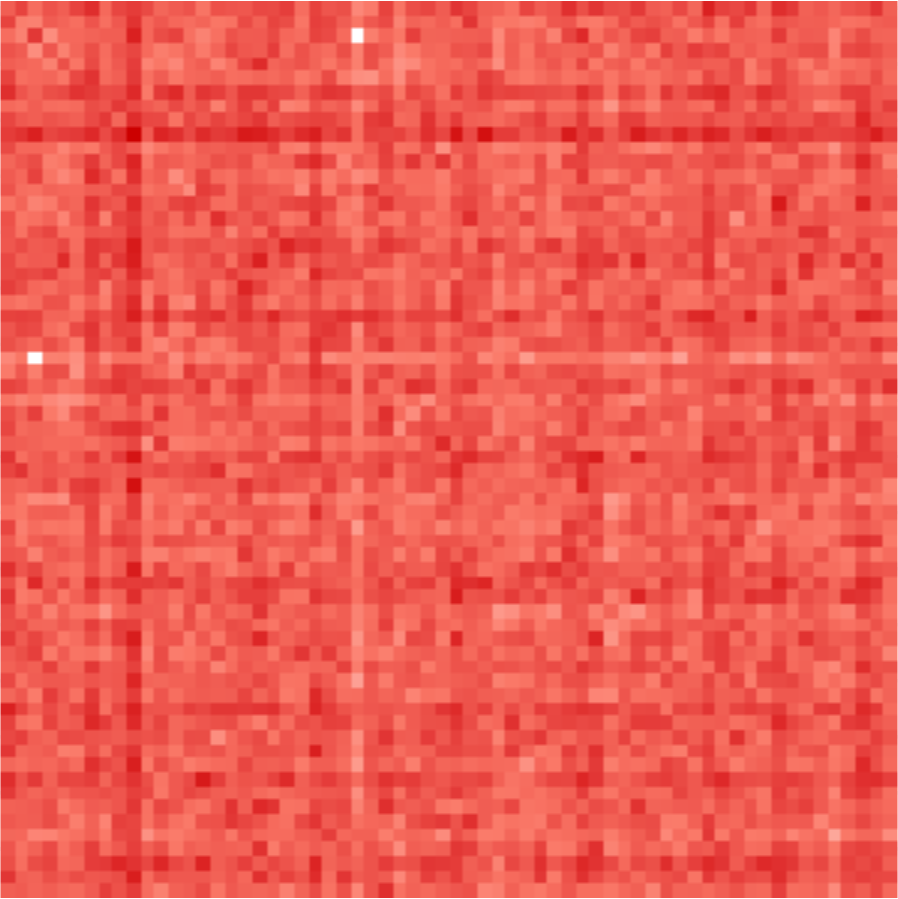}{Random} &
    \ordhead{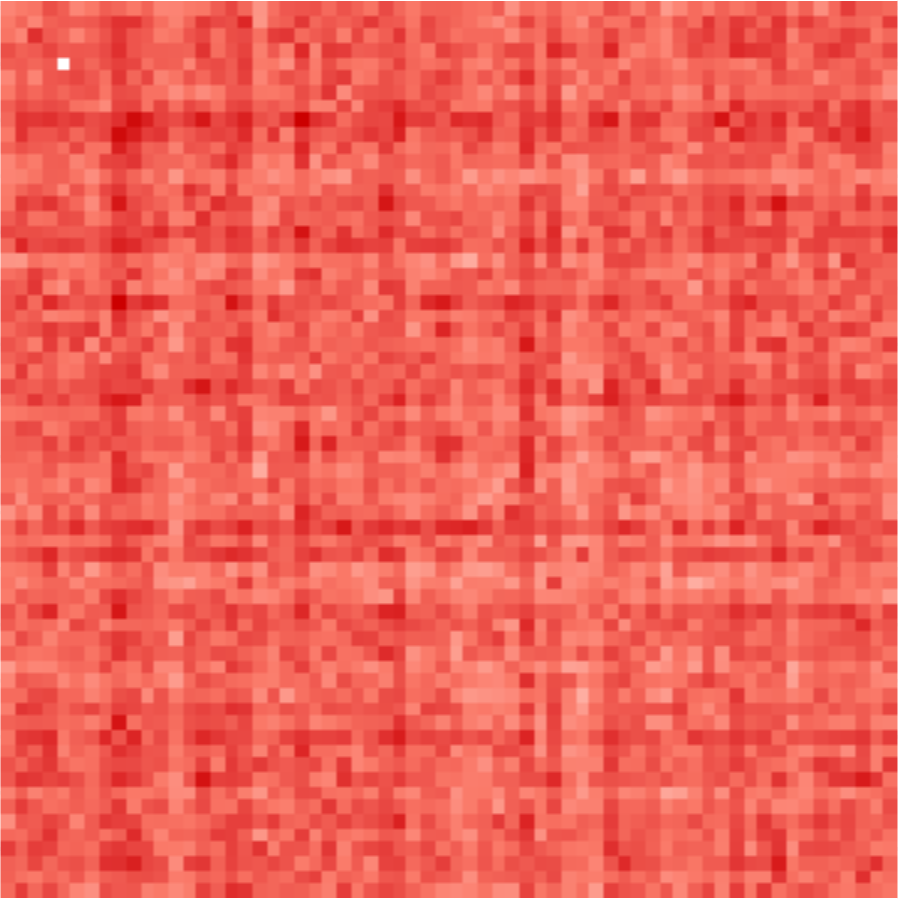}{Gorder} &
    \ordhead{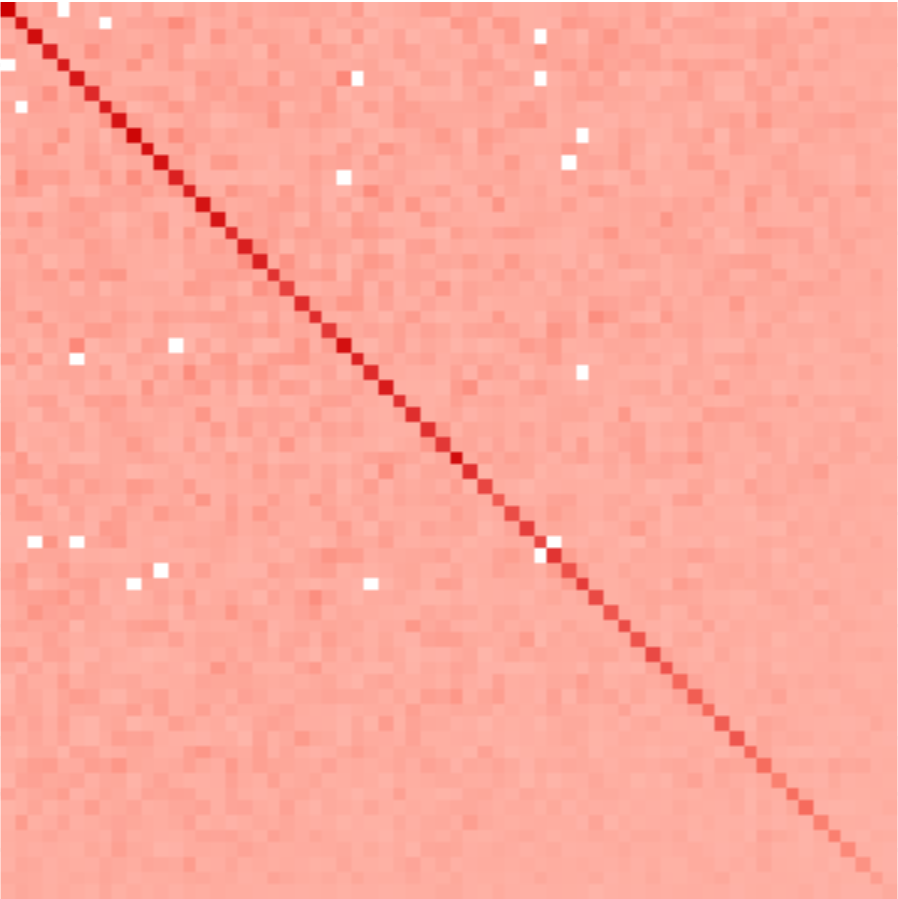}{\begin{tabular}[c]{@{}c@{}}Iter.\\ Clust.\end{tabular}} &
    \ordhead{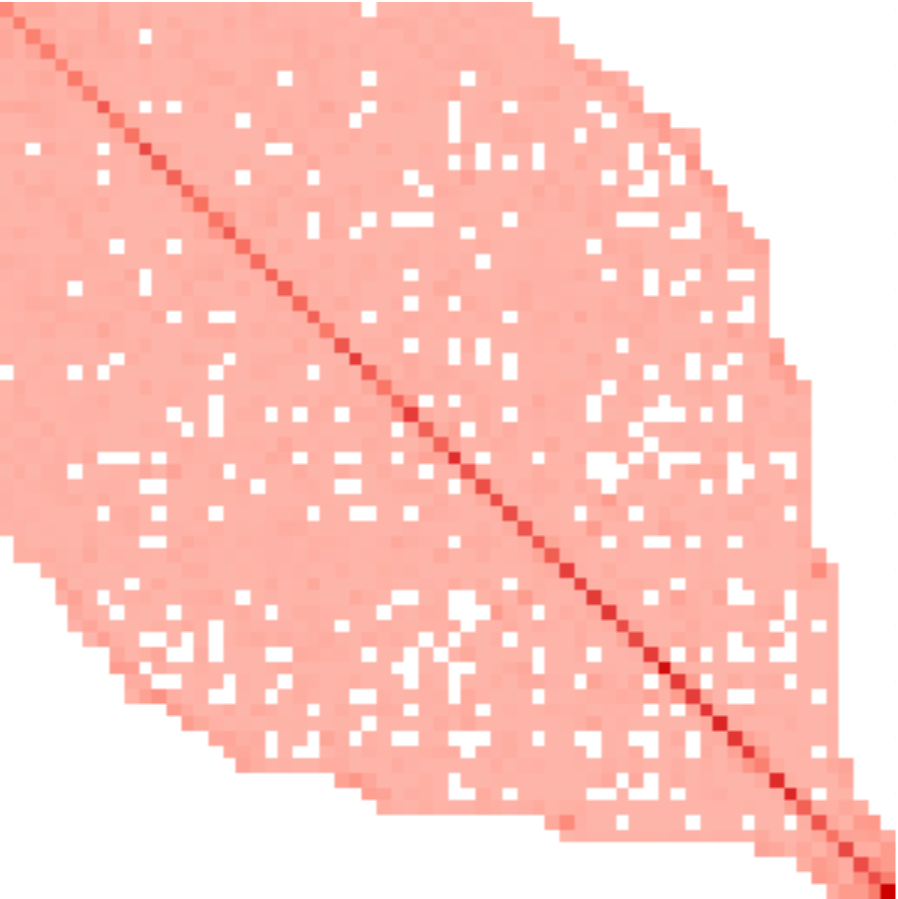}{RCM} &
    \ordhead{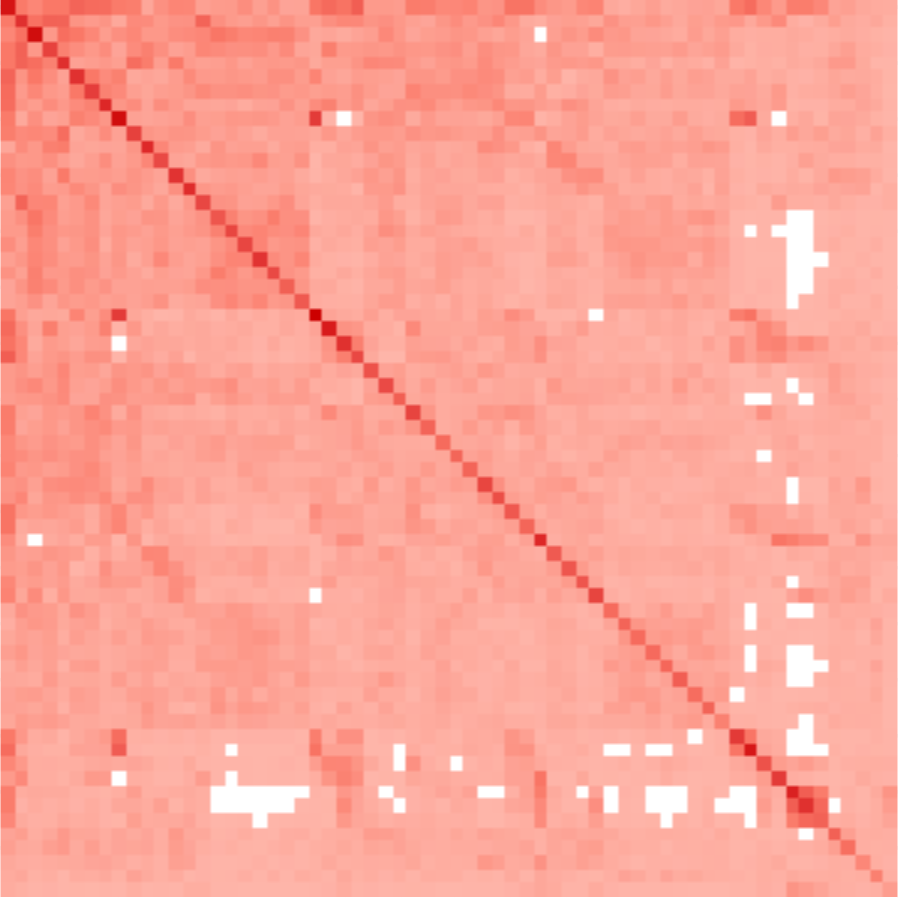}{\begin{tabular}[c]{@{}c@{}}Proposed\\ (Alg.~\ref{alg:jaccard-window})\end{tabular}} \\
    \hline
    \textbf{\begin{tabular}[c]{@{}l@{}}Comp. ratio\end{tabular}} 
    & 12\% & 12\% & 48\% & 46\% &   73\% \\ 
    \textbf{\begin{tabular}[l]{@{}l@{}}Order. time\end{tabular}}
    & 0.06 & 1.19 & 21.69 & 0.25 &  15.49 \\
    \hline
  \end{tabular}
  }
  \caption{\small
    Compression ratio and preprocessing time for {\tt{\footnotesize vsp\_msc}}.
  }
  \label{tab:compression_ordering}
\end{subtable}
  \hfill
  \begin{subtable}[t]{0.4\linewidth}
    \centering
    \scalebox{1.3}{
    \begin{tabular}{lrr}\\  \\
      \textbf{Graph} &
      \textbf{\begin{tabular}[r]{@{}r@{}}${\mathcal{U}_{div}}$\\(unordered)\end{tabular}} &
      \textbf{\begin{tabular}[r]{@{}r@{}}${\mathcal{U}_{div}}$\\(ordered)\end{tabular}} \\
      \hline
      GAP-road      & 158696  & 869  \\
      europe\_osm   & 869302  & 982  \\
      delaunay\_n24 & 894542  & 2839 \\
      rgg\_24       & 3948    & 1512 \\
      \hline 
    \end{tabular}
    }
    \caption{\small
      Average update divergence ${\mathcal{U}_{div}}$ before/after ordering.
    }
    \label{tab:locality_values}
  \end{subtable}

  \caption{\small
    Effect of graph ordering \acro's BVSS:
    (a) The compression ratio and preprocessing time for {\tt{vsp\_msc}}~\cite{suitesparse} with 22K vertices and 2.4M edges having different orderings, including {\sc{JaccardWithWindows}}~(Algorithm~\ref{alg:jaccard-window}).
    (b) The reduction on update divergence $\mathcal{U}_{div}$ for non-social networks before and after applying RCM.
  }
  \label{tab:ordering_overview}
\end{table*}

Having a higher average number of set bits in slices is useful to reduce (1) the memory footprint of BVSS, and (2) the time/power spent on unnecessary checks for non-existing incoming edges. Recall that the connectivity pattern of a slice with row ID $u$ in slice set $s$ is constructed from the entries $\mathbf{A}[u][\sigma s + i]$ for $0 \le i < \sigma$, i.e., from a contiguous block of $\sigma$ columns. Let $m$ be the number of edges in the graph $G = (\mathcal{V},\mathcal{E})$~(or the nonzeros in the adjacency matrix $\mathbf{A}$), and let \textit{numSlices} be the number of \emph{non-empty (unpadded)} slices. Since each slice can encode $\sigma$ edges, the effective {\em{compression ratio}} is $\frac{m}{\textit{numSlices} \times \sigma}$, which is the fraction of set bits in masks. To compress more, one can reorder $G$ so that vertices with common outgoing neighbours are placed in the same slice sets, thereby packing more edges into fewer slices. This problem is closely related to the community detection problem and has been extensively studied in the literature~\cite{gorder, rabbit, rabbit++, sylos_labini, rcm, spmv_blocking}.

Let $N(u)$ and $N(v)$ denote the adjacency lists of $u, v\in \mathcal{V}$. There exist ordering heuristics in the literature which place $u$ and $v$ close based on their {\em Jaccard~ similarity} $J(u,v) = \frac{\lvert N(u) \cap N(v) \rvert}{\lvert N(u) \cup N(v) \rvert}$~\cite{jaccard}. A high Jaccard similarity~(i.e., closer to $1$) implies the existence of many common neighbours, and taking it into account helps to reduce the number of mostly empty slices.

Various, similarity-focused heuristics have been proposed to order a graph, including {\em{Gorder}}~\cite{gorder}, {\em Rabbit}~\cite{rabbit}, and {\em IterativeClustering}~\cite{sylos_labini}. {\em RCM}~\cite{rcm}, on the other hand, does not explicitly target communities, but it still generates slice sets desirable in \acro's BVSS by putting nonzeros close to the diagonal, which is especially useful for graphs with high diameters. Among those stated, only {\em IterativeClustering}~\cite{sylos_labini} uses the {\em Jaccard} as its core metric. It simply builds vertex clusters by greedily adding the most similar vertex to the current cluster being constructed. If there is no {\em good} vertex, it starts a new cluster. The process continues until all vertices are selected. In our preliminary experiments, this heuristic generated the best orderings. Unfortunately, it is too expensive for our purposes.

In this work, we propose a {\em hybrid} ordering heuristic that exploits the fact that only a small number, $\sigma$, of columns (which will form a slice set) must exhibit high intra-group Jaccard similarity. Let $w \ll n$ be a multiple of $\sigma$. To reduce the cost, \acro divides the columns into $\lceil n / w \rceil$ windows of size $w$~(wlog., assume that $w$ divides $n$) and runs an independent Jaccard-based clustering inside each window. Within a single window, \acro creates exactly $w / \sigma$ clusters, each corresponding to a slice set. When $w$ is sufficiently large, many of the clusters generated should have a nice compression ratio. 

To generate a cluster, we start from an arbitrary seed vertex and iteratively select the most similar $\sigma - 1$ vertices from the same window. In short, we perform $\lceil n/w \rceil$ Jaccard-based, greedy orderings, as given in Algorithm~\ref{alg:jaccard-window}, each restricted to the $w$ vertices in a separate window. As expected, although this restriction reduces the runtime, it also reduces the compression ratio since the windows may not contain good cluster candidates. To boost the heuristic's effectiveness, we first apply Gorder~\cite{gorder} and increase the likelihood of having Jaccard-wise similar vertices in the same window, thereby improving the compression ratio even with smaller values of $w$.

Since the windows are independent, this hybrid heuristic is pleasingly parallel. A trivial implementation performs $\mathcal{O}(w^2 \times \delta)$ work per window, where $\delta$ is the maximum degree in $G$. Since there are $n / w$ windows, the overall complexity becomes $\mathcal{O}(n \times w \times \delta)$, which is feasible for moderate $w \ll n$. We compare the heuristic's compression performance, when $w = 2^{16}$, against the aforementioned reordering algorithms applied to the BVSS, for a random star graph {\tt{\small vsp\_msc}}, taken from SuiteSparse~\cite{suitesparse}, in Table~\ref{tab:compression_ordering}.

\begin{algorithm}[t]
\renewcommand{\baselinestretch}{0.93}
\caption{: \sc{JaccardWithWindows}}
\label{alg:jaccard-window}
\small
\begin{algorithmic}[1]
\Require {(1) $\mathbf{A} \in \{0,1\}^{n \times n}$ in CSC, 
(2) $\sigma$: slice size, (3) $w$: window size with $w \equiv 0 \pmod{\sigma}$}
\Ensure $\pi^{-1} : \{0,\dots,n-1\} \to \{0,\dots,n-1\}$: inv. permutation for $\mathcal{V}$

\State Partition $\{0,\dots,n-1\}$ into disjoint windows $W_k = [s_k, e_k)$ of size $\le w$
\ForAll{$W_k$ \textbf{in parallel}}
  \State $Q \gets \{j : j \in W_k\}$
  \State $L_k \gets e_k - s_k$  \Comment{window length}
  \State $M_k \gets \lceil L_k / \sigma \rceil$  \Comment{number of clusters (slice sets)}
  \For{$\ell = 0,\dots,M_k-1$}
    \State $j^* \gets \mathrm{pop}(Q)$ \Comment{first, singleton vertex}
    \State $\pi^{-1}(j^*) \gets s_k + \ell \cdot \sigma$ \Comment{global position}
    \State $U \gets N(j^*)$
    \For{$r = 1,\dots,\sigma-1$} \Comment{for each position inside cluster}
      \If{$Q = \emptyset$}
        \State \textbf{break}
      \EndIf
      \State $j^\dagger \gets \displaystyle\arg\max_{j \in Q} \frac{|N(j)\cap U|}{|N(j)\cup U|}$
      \State Remove $j^\dagger$ from $Q$
      \State $\pi^{-1}(j^\dagger) \gets s_k + \ell \cdot \sigma + r$
      \State $U \gets U \cup N(j^\dagger)$
    \EndFor
  \EndFor
\EndFor
\State \Return $\pi^{-1}$
\end{algorithmic}
\renewcommand{\baselinestretch}{1}
\end{algorithm}

\subsubsection{Reducing excessive cache line fetch}
\label{sec:ordering_cache_line}

Although increasing the compression ratio is crucial, an equally important objective is to reduce the number of excessive cache-line fetches during \acro's update phase. Although social networks have communities, for non-social, i.e., road or geometric, networks, we observe that VSSs already tend to have fewer slices, so compressing them yields little additional benefit. In contrast, the updates remain expensive because the number of cache lines touched is directly related to the {\em spread} of row IDs within each VSS. This is why \acro applies a different, simpler ordering strategy for non-social networks.

To quantify the spread, we define the \emph{update divergence} metric. For each of the $slicesPerThread$ columns of the slide-to-thread distribution matrix, we denote the {\em{column divergence}} as the {\em{ standard deviation}} of the pulling row IDs with nonzero masks~(i.e., padded slices are discarded). Then the {\em set divergence} for a VSS $s$, $\mathcal{U}_{div}(s)$, is the average of its column divergences over non-empty columns. Finally, the {\em{update divergence}} is the average $\mathcal{U}_{div} = \underset{\{s: s\ \textnormal{is a VSS} \}}{\tt{avg}}\{\mathcal{U}_{div}(s)\}$ computed over all VSSs.

On Nvidia GPUs, global memory coherence is maintained at L2 cache shared across all SMs, whereas each SM has its private L1. Writes (including atomics) to global memory are serviced through L2 and typically use a write-through policy from L1 to L2. Hence, every store forces the corresponding 128-byte cache line to be updated in L2 and may invalidate/bypass any copies in L1. This means that if a cache line initially resides only in GPU DRAM, it must be fetched into L2 before the atomic can complete; if multiple SMs frequently write to the same line, that line is repeatedly written through to L2 and re-fetched into whichever L1 cache needs it next.

As illustrated in Fig.~\ref{fig:data_structure}, the next frontier is atomically updated based on the ID of row $u$, i.e., via an index \textbf{F}$_{\text{next}}$$[\lfloor u / \sigma \rfloor]$. Reducing the ID range within a VSS directly reduces the number of distinct 128-byte L2 lines (backing up \textbf{F}$_{\text{next}}$) that a warp fetches from DRAM. The second update indexed by the $u$ is on the \textbf{\textit{level}} array (omitted in Fig.~\ref{fig:data_structure} for simplicity). Although this update does not incur atomics, narrowing the range of vertex IDs within a VSS still brings substantial benefit: non-atomic loads and stores to the \textbf{\textit{level}} array can hit in the per-SM L1 cache, and when a warp updates vertices clustered in a narrow region, i.e., with a lower {\em divergence}, it is much more likely that the required cache lines are already present in L1.

To reduce $\mathcal{U}_{div}$ on a non-social network, we apply a widely-used bandwidth-reducing ordering, RCM~\cite{rcm}. A smaller bandwidth implies that the number of cache lines that must be fetched (either from DRAM for frontier updates or from L2 for level updates) during the update phase is reduced. We observed that when RCM is effective, i.e., when there is a noticeable reduction in bandwidth, performance improves. Table~\ref{tab:locality_values} reports the update divergences with and without RCM on non-social networks. 

To decide the ordering strategy, \acro assumes a network is {\em social-like}~(or {\em scale-free-like}) if it has at least one of these properties: (i) \emph{heavy-tail sharing}, and (ii) approximate \emph{power-law behaviour}. \acro assumes that when the top-$1\%$ and top-$10\%$ of the vertices w.r.t. to the degree has more than $5\%$ and $40\%$, respectively, of the edges, the degree distribution in $G$ is heavy-tailed. Similarly, if the log–log transform of the degree histogram fits to a straight line with a slope typical for scale-free networks, the network is assumed to show a power-law behaviour. If $G$ does not have either of these properties, \acro uses RCM to reorder $G$. 

Since social-like networks inherently have high bandwidth, even when applied, RCM may fail to reduce their update divergence in a meaningful way. Furthermore, for large networks, its applicability can be limited by its $\mathcal{O}(m \delta)$ time complexity~\cite{rcm_complexity}, where $\delta$ is the maximum degree in $G$ and is usually large.

\section{Compute Mechanics of \acro}
\label{sec:algo}
Although the BVSS data structure already addresses several major issues of the state-of-the-art Tensor-Core-based BFS implementations—namely load imbalance, frontier-oblivious slice set distribution, expensive vertex updates, and unsatisfactory compression ratios—the \emph{algorithmic} side of the pipeline, i.e., how TCs are actually used, and the way that the pull updates are performed, still leaves considerable room for improvement.

\subsection{An Optimal Layout for TC Multiplication}
\label{sec:mult_pattern}
Let us assume $\sigma=8$ in this discussion. To the best of our knowledge, the literature suggests using 16 consecutive TC calls for the $\sigma$-bit pulls from 128 vertices~\cite{berrybees}. The state-of-the-art arranges 128 slices of a slice set by the $32 \times 4$ slice-to-thread distribution matrix, as illustrated in part~(b) of Fig.~\ref{fig:computation_structure}. This distribution matrix is processed in four \emph{rounds}; in each round, a single column of the $32 \times 4$ distribution matrix is selected and reorganised into an $8 \times 4$ \textit{fragA} matrix. The processing of a column is split into four \emph{periods}. In each period, a block of 8 rows is fed into \textit{fragA} and only a single column of the $4 \times 8$ \textit{fragB} matrix is effectively used; the active column index $j \in \{0,1,2,3\}$ coincides with the current period ID. As a result, each TC call produces popcounts for only 8 dot products, and only the $\sigma$-bit portion of a single 32-bit word in \textit{fragB} is actually used, turning the matrix–matrix primitive into a series of narrow matrix–vector multiplications. Since there are four rounds (one per column) and four periods per round, a total of $4 \times 4 = 16$ TC multiplications are required to process 128 slices assigned to a warp.

\acro uses TC cores in an optimal fashion, i.e., without redundancy; the $8 \times 8$ popcount output matrix has 64 distinct entries, each capable of encoding the result of one dot product. Thus, in principle, a single TC call can resolve 64 independent dot products. For a VSS with 128 slices, 128 dot products are required. Hence, any data layout and multiplication pattern requires at least two TC calls.
We propose a novel multiplication pattern that converts the matrix–matrix multiplication primitive into a batched matrix–vector multiplication tailored to the BVSS layout of \acro. The pattern packs the computation so that all 64 entries of the $8 \times 8$ popcount output are useful (i.e., no output is wasted), resolving 64 dot products per TC operation. Consequently, \acro requires only two TC multiplications per VSS, 1/8 of the calls needed by the SotA. The overall pipeline is depicted in part~(c) of Fig.~\ref{fig:computation_structure}.

\algnewcommand{\IIf}[1]{\State\algorithmicif\ #1\ \algorithmicthen}
\algnewcommand{\EndIIf}{\unskip\ }
\algrenewcommand\algorithmicindent{0.9em}%

\begin{algorithm}[t]
\small
\renewcommand{\baselinestretch}{0.93}
\caption{\acro{}}
\label{alg:bvss_kernel}
\begin{algorithmic}[1]
\Require (1) BVSS data structure, (2) $src$: source vertex
\Ensure (1) \(\textbf{L}\): level array

\State \(\textbf{L}[v] \gets \infty,\ \forall v \in \mathcal{V} /\{src\}\); \(\textbf{L}[src] \gets 0\); \Comment{initialize} 
\State \(\textbf{{F}}_{\text{curr}}[src] \gets 1\); \Comment{initialize}
\State \(\ell \gets 0;\) \Comment{current BFS level}
\State \(\textit{cont} \gets \texttt{\footnotesize true};\) \Comment{non-empty state of the frontier}

\While{\(\textit{cont}\)} 
  \State \(\ell \gets \ell + 1;\)
  \For{\(w = \textit{warpID};\ w < |\textbf{Q}_{\text{curr}}|;\ w \gets w + \#\text{warps}\)}
    \State \(vss_{in} \gets \textbf{Q}_{\text{curr}}[w];\)        \Comment{virtual slice set ID}
    \State \(ss_{in} \gets \textit{\textbf{virtualToReal}}[vss_{in}];\) \Comment{slice set ID}
    \State \(\alpha \gets \textbf{{F}}_{\text{curr}}[ss_{in}];\)   \Comment{$\sigma$-bit frontier word}

    \State \(\textit{tile} \gets (vss_{in} \ll 5) + \textit{laneID};\)
    \State \((u_0,u_1,u_2,u_3) \gets \textit{\textbf{rowIds}}[\textit{tile}];\) \Comment{vectorized 128-bit read}
    \State \(\textit{mask} \gets \textit{masks}[\textit{tile}];\) \Comment{4 x $\sigma$ bit read}

    \State \(\textit{fragB} \gets 0;\)
    \IIf{\(\textit{laneID} \bmod 9 = 0\)}      
    \(\textit{fragB} \gets \alpha;\)
    \EndIIf
    \IIf{\(\textit{laneID} \bmod 9 = 4\)}      
       \(\textit{fragB} \gets \alpha;\)
       \(\textit{fragB} \gets \textit{fragB} \ll 8;\) 
    \EndIIf

    \State \(\textit{fragA} \gets \text{low16}(\textit{mask});\) \Comment{the first two masks}
    \State \(\textit{fragC}[0,1] \gets \mathrm{TC}(\textit{fragA}, \textit{fragB});\) \Comment{1st m8n8k128}

    \State \(\textit{fragA} \gets \text{high16}(\textit{mask});\) \Comment{the last two masks}
    \State \(\textit{fragC}[2,3] \gets \mathrm{TC}(\textit{fragA}, \textit{fragB});\) \Comment{2nd m8n8k128}

    \For{\(c \in \{0,1,2,3\}\)} \Comment{column in slice distribution}
      \If{\(\textit{fragC}[c] \neq 0\)} \Comment{if dot-product is nonzero}
        \State \(u \gets (u_0,u_1,u_2,u_3)[c];\) \Comment{row ID to update}
        \State \(\ell_{\text{prev}} \gets \textbf{L}[u];\)

        \If{\(\ell < \ell_{\text{prev}}\)} \Comment{if $u$ is not visited}
          \State \(\textbf{L}[u] \gets \ell;\) \Comment{set $u$'s level}
          \State \(old \gets \textbf{{F}}_{\text{next}}[u] \stackrel{{\text{atomic}}}{\lor} 1;\)
          \If{\(old = 0\)} \Comment{$u$ is not in ${\textbf{F}}_{\text{next}}$}
            \State \({ss_{out}} \gets \lfloor u / \sigma \rfloor;\) \Comment{$u$'s slice set index}
            \State \([start,end) \gets [\textit{\textbf{realPtrs}}[ss_{out}],\ \textit{\textbf{realPtrs}}[ss_{out}+1]);\)
            \State \(\textbf{Q}_{\text{next}} \;\stackrel{\text{atomic}}{\gets}\; \textbf{Q}_{\text{next}} \;\cup\; [start,end);\)
          \EndIf
        \EndIf
      \EndIf
    \EndFor
  \EndFor

  \State \textsc{GridSync}(); \Comment{level synchronization}
  \State \(\textit{cont} \gets (|\textbf{Q}_{\text{next}}| > 0);\) \Comment{check the frontier state} 
  \State \(\text{swap}({\textnormal{\textbf{F}}}_{\text{curr}}, {\textnormal{\textbf{F}}}_{\text{next}});\) \(\text{swap}(\textbf{Q}_{\text{curr}}, \textbf{Q}_{\text{next}});\) \Comment{swap for the next level}
  \State \textsc{GridSync}();
  \State \(|\textbf{Q}_{\text{next}}| \gets 0;\) \({\textnormal{\textbf{F}}}_{\text{next}} \gets 0;\) \Comment{clear the next frontier data}
  \State \textsc{GridSync}();
\EndWhile
\end{algorithmic}
\renewcommand{\baselinestretch}{1}
\end{algorithm}

In each round, \acro processes two of the four columns of the $32 \times 4$ distribution matrix. These columns form a $32 \times 2$ intermediate matrix, which we conceptually reshape into an $8 \times 8$ intermediate \textit{fragA} tile~(shown with dashed lines). Since the TC instruction expects \textit{fragA} to have shape $8 \times 4$, we pack two slices into each \textit{fragA} entry: for every row in the $8 \times 8$ intermediate tile, the two $\sigma$-bit connectivity patterns belonging to the two selected columns are concatenated into a single $2\sigma$-bit field inside a 32-bit \textit{fragA} element. To effectively and efficiently utilise all \textit{fragC} entries, \acro uses a single $4 \times 8$ \textit{fragB} shared across all rounds. Let the $\sigma$-bit frontier word be $\alpha$ and the all-zero $\sigma$-bit word be $\theta$. We define 16-bit building blocks $X = \alpha\theta$, $Y = \theta\alpha$, and $Z = \theta\theta$. Each row $i \in \{0,1,2,3\}$ of \textit{fragB} contains exactly two nonzero 32-bit words at columns $j = 2i$ and $j = 2i + 1$. The other \textit{fragB} entries are all-zero words. For the eight non-zero entries, if the column index $j$ is even, the corresponding 32-bit word is $XZ$, and if $j$ is odd, it is $YZ$. Hence, for even $j$, \textit{fragB} selects the first packed slice (via $X$), and for odd $j$, it selects the second packed slice (via $Y$), while the $Z$s mask out the unused halves. To simplify the accumulation of dot-product results, we exploit the distribution of \textit{fragB} words to the warp threads, as shown in part~(a) of Fig.~\ref{fig:computation_structure}. The threads $t$ with $\textit{t} \bmod 9 = 0$ hold the 32-bit word $XZ$ in their \textit{fragB} registers, while threads with $\textit{t} \bmod 9 = 4$ hold the word $YZ$. All remaining threads hold $0$ in \textit{fragB} registers. 

The layouts of \textit{fragA} and \textit{fragB} not only make each \textit{fragC} entry useful, but also ensure that each thread gets the results of its own slice–frontier dot products directly from its \textit{fragC} registers, without extra intra-warp communication. Note that each thread $t$ holds the \textit{fragC} words located at $(i,j)$ and $(i,j+1)$ with $i = \lfloor t/4\rfloor$ and $j = 2 \times (t \bmod 4)$. This happens because $t$’s slices are packed in \textit{fragA}[$i$][$\frac{j}{2}$]~(Fig.~\ref{fig:computation_structure}.a). This \textit{fragA} word is multiplied by \textit{fragB}[$\frac{j}{2}$][$j$] $= XZ$ and \textit{fragB}[$\frac{j}{2}$][$j + 1$] $= YZ$ which select the first and second slice masks, respectively, of $t$, and the outputs are written to \textit{fragC}[$i$][$j$] and \textit{fragC}[$i$][$j + 1$] owned by $t$. Hence, with the proposed data layout and multiplication pattern, \acro ensures that each output of the popcount matrix is useful and stored where it will be needed. The \acro algorithm is summarised in Algorithm~\ref{alg:bvss_kernel}.

\subsection{Lazy Vertex Updates}
\label{sec:lazy_vertex}
In Algorithm~\ref{alg:bvss_kernel}, each thread performs two atomic operations per pull. In particular, it is expected for {\em level} checks/updates (lines 24--26) to exhibit a low L1 hit rate, while {\em frontier} checks/updates (lines 27--31) suffer from a low L2 hit rate, as detailed in Section~\ref{sec:ordering_cache_line}. To address this, we propose a \emph{lazy vertex update} scheme that defers updates to the end of each BFS level and applies them in a fully coalesced manner by sweeping {\textnormal{\textbf{F}}}$_{\text{next}}$ without being affected by the high update divergence. This not only mitigates the cache-hit problems but also substantially alleviates the cost of atomics.

The Instruction Set Architecture (ISA) of GPUs, called SASS, supports two types of atomic instructions, although the high-level language, from which it is compiled, abstracts this complexity away. These instructions are \textbf{ATOMG} and \textbf{REDG}, both of which guarantee atomicity, with the latter being the asynchronous version of the former. The decision as to which of them the high-level code compiles to is determined by the backend compiler. If the code relies on the return value of an atomic\footnote{All CUDA atomics are capable of returning the value before it is modified.}, then the compiler emits an \textbf{ATOMG} instruction, a full atomic that stalls the requesting warp until the return value is written back to registers. If, on the other hand, the code ignores the return value of the atomic, the compiler can use the lighter-weight \textbf{REDG} instruction, which performs the update asynchronously and allows the warp to continue execution without waiting for the result.

\algrenewcommand\algorithmicindent{1.1em}%

\begin{algorithm}[t]
\small
\renewcommand{\baselinestretch}{0.93}
\caption{\acro{} with Lazy Vertex Updates}
\label{alg:bvss_kernel_lazy}
\begin{algorithmic}[1]
\Require (1) BVSS data structure, (2) $src$: source vertex
\Ensure (1) \(\textbf{L}\): level array

\State \(\textbf{L}[v] \gets \infty,\ \forall v \in \mathcal{V}/\{src\}\); \(\textbf{L}[src] \gets 0\); \Comment{initialize}
\State \({\textnormal{\textbf{F}}}_{\text{curr}}[src] \gets {\textnormal{\textbf{V}}}_{\text{curr}}[src] \gets {\textnormal{\textbf{V}}}_{\text{next}}[src] \gets 1\); \Comment{initialize}
\State \(\ell \gets 0;\) \Comment{current BFS level}
\State \(\textit{cont} \gets \texttt{true};\) \Comment{non-empty state of the frontier}

\While{\(\textit{cont}\)}
  \State \(\ell \gets \ell + 1;\)

  \State \textbf{Stage 1: Lazy marking\ \hrulefill}
  \For{\(w = \textit{warpID};\ w < \lvert \textbf{Q}_{\text{curr}} \rvert;\ w \gets w + \#\text{warps}\)}
    \State ... \\ \hspace*{7.1ex}Lines 8--20 of Algorithm~\ref{alg:bvss_kernel}\\\hspace*{7.1ex}...
      \setcounter{ALG@line}{25}

    \For{\(c \in \{0,1,2,3\}\)}\Comment{column in slice distribution}
      \If{\(\textit{fragC}[c] \neq 0\)} \Comment{if dot-product is nonzero}
        \State \(u \gets (u_0,u_1,u_2,u_3)[c];\) \Comment{row ID to update}
        \State \({\textnormal{\textbf{V}}}_{\text{next}}[u] \;\stackrel{\text{atomic}}{\lor}\; 1;\) \Comment{lazy mark~(no return value use)}
      \EndIf
    \EndFor
  \EndFor

  \State \(\lvert \textbf{Q}_{\text{next}} \rvert \gets 0;\)\Comment{prepare for Stage 2}
  \State \textsc{GridSync}();
   \vspace*{1ex}
  \State \textbf{Stage 2: Frontier Finalization\ \hrulefill} 
   \vspace*{0.7ex}
  \For{\(t = \textit{threadID};\ t < \#\text{words};\ t \gets t + \#\text{threads}\)}
    \State \(\textit{next} \gets {\textnormal{\textbf{V}}}_{\text{next}}[t];\)\Comment{are vertices visited until here}
    \State \(\textit{diff} \gets {\textnormal{\textbf{V}}}_{\text{curr}}[t] \,\mathbin{\underline{\vee}}\, \textit{next};\) \Comment{is something changed in Stage 1}
    \State \(\textit{rssOffset} \gets 4t;\) \Comment{4 slice sets per word}
    \If{\(\textit{diff} \neq 0\)} \Comment{a vertex in this word is in the next frontier }
      \State \({\textnormal{\textbf{V}}}_{\text{curr}}[t] \gets \textit{next};\)
      \State \({\textnormal{\textbf{F}}}_{\text{curr}}[t] \gets \textit{diff};\) \Comment{set the bits for frontier vertices}
      \For{\(\textit{set} \in \{0,1,2,3\}\)} \Comment{for each 8-bit words in $diff$}
        \State \(ss_{mask} \gets (\textit{diff} \gg (8 \cdot \textit{set})) \mathbin{\&}\ 0x\text{FF};\) \Comment{select 8-bits}
        \If{\(ss_{mask}  \neq 0\)} \Comment{if the change is in this set}
          \State \(\textit{$ss_{out}$} \gets \textit{rssOffset} + \textit{set};\)
          \While{\(ss_{mask}  \neq 0\)} \Comment{go over all the bits}
            \State \(b \gets \textbf{\_ffs}(ss_{mask} ) - 1;\) \Comment{in \(\{0,\dots,7\}\)}
            \State \(u \gets ss_{out} \cdot \sigma + b;\) \Comment{a new frontier vertex}
            \State \(\textbf{L}[u] \gets \ell;\) 
            \State \(ss_{mask}  \gets ss_{mask}  \mathbin{\&} (ss_{mask}  - 1);\) \Comment{unset $u$'s bit}
          \EndWhile
            \State \([start,end) \gets [\textit{\textbf{realPtrs}}[ss_{out}],\ \textit{\textbf{realPtrs}}[ss_{out}+1]);\)
          \State \(\textbf{Q}_{\text{next}} \;\stackrel{\text{warp-atomic}}{\gets}\; \textbf{Q}_{\text{next}} \;\cup\; [start,end);\)
        \EndIf
      \EndFor
    \EndIf
  \EndFor

  \State \textsc{GridSync}();
  \State \(\textit{cont} \gets (\lvert \textbf{Q}_{\text{next}} \rvert > 0);\)
  \State \(\text{swap}(\textbf{Q}_{\text{curr}}, \textbf{Q}_{\text{next}});\) \(\text{swap}(\lvert \textbf{Q}_{\text{curr}} \rvert, \lvert \textbf{Q}_{\text{next}} \rvert);\)\Comment{swap for the next $\ell$}
  \State \textsc{GridSync}();
\EndWhile
\end{algorithmic}
\renewcommand{\baselinestretch}{1}
\end{algorithm}

As seen in line 27 of Alg.~\ref{alg:bvss_kernel}, the current kernel relies on the result of the atomic~(line 28) to decide whether to push the VSSs of the slice set in which the pulling vertex resides. In the lazy update scheme, \acro defers this decision to the end of the BFS level and instead marks vertices lazily, using the lightweight asynchronous atomic operation. At the end of each level, once all lazy marking is finished, \acro examines all vertices and inserts the corresponding VSSs of those that are marked into the queue, without introducing any full-atomic instruction. This approach, while relying on a \(\Theta(n)\) loop per BFS level, resolves the problems discussed above.

\begin{table*}[t]
\centering
\scalebox{0.90}{
\setlength{\tabcolsep}{2.6pt}
\begin{tabular}{c|lc||rr||rrrr||r|rr|r|r}
\multicolumn{5}{c}{}& \multicolumn{8}{c}{\textbf{Times (in milliseconds)}}  \\ \cline{6-13}
& & {\bf{Soc.}} & & & \multicolumn{1}{c}{\textbf{GAP}}  & \multicolumn{1}{c}{\textbf{Gunrock}} & \multicolumn{1}{c}{\textbf{GSWITCH}}  & \multicolumn{1}{l||}{\textbf{Berry}} & \multicolumn{4}{c|}{\textbf{\acro}} & \textbf{\acro}  \\
& \textbf{Graph}                      &
{\bf{Like}} & 
\textbf{$|\mathcal{V}|$}              &
\textbf{$|\mathcal{E}|$}              &
\multicolumn{1}{c}{\cite{gap}}        &
\multicolumn{1}{c}{\cite{Wang:2017:GGG}}                        &
\multicolumn{1}{c}{\cite{gswitch}}    &
\textbf{Bees}~\cite{berrybees}   &
\multicolumn{1}{c|}{\textbf{(a)}}                    &
\multicolumn{1}{c}{\textbf{(ac)}}                   &
\multicolumn{1}{c|}{\textbf{(ab)}}                   &
\multicolumn{1}{c|}{\textbf{(full)}}                  &
vs.~\cite{berrybees}        \\
\midrule
\multirow{5}{*}{\rotatebox[origin=c]{90}{GAP}}
&{\tt{GAP-road}} & \text{\sffamily X}            & 23M & 57M & 739.64 & 317.53 & 130.20 & 161.92     & 49.09         & {{{73.08}}}         & 41.23         & 41.23     & 3.93$\times$ \\
&{\tt{GAP-twitter}} & $\checkmark$         & 61M & 1.4B & 201.54 & (error) & (error) & 63.66    & 54.51         & 20.56         & 45.22         & 19.21     & 3.31$\times$ \\
&{\tt{GAP-web}} &  $\checkmark$               & 50M & 1.9B & 327.14 & (error) & 277.80 &  74.25  & 16.86         & {{{25.28}}}         & {\bf{{27.30}}}         & 16.32     & 4.55$\times$ \\
&{\tt{GAP-kron}} &  $\checkmark$             & 134M & 4.2B & 290.15 & (error) & (error) & 164.93  & 150.97   & 57.58    & 149.98 & 57.54     & 2.87$\times$  \\
&{\tt{GAP-urand}} &  $\checkmark$             & 134M & 4.2B & 468.82 & (error) & (error) &  192.90 & 201.89        & 52.24         & 200.02        & 52.19     & 3.70$\times$ \\
\midrule
\multirow{10}{*}{\rotatebox[origin=c]{90}{$|\mathcal{V}| \ge 23M$, $|\mathcal{E}| \le 2^{32} - 1$}}
&{\tt{nlpkkt240}} &  $\checkmark$             & 27M & 760M & 234.26 & 25.12 & 66.44 & 33.32      & 11.16         & 11.22         & {\bf{{17.41}}}         & {\textit{\textbf{15.85}}}    & 2.10$\times$ \\
&{\tt{uk-2005}} &  $\checkmark$                & 39M & 936M & 177.19 & 331.75 & 36.25 & 22.78     & 11.59         & {{{14.77}}}         & {\bf{{21.42}}}         & 10.83     & 2.10$\times$ \\
&{\tt{it-2004}} &  $\checkmark$                & 41M & 1.1B & 216.22 & (error) & 97.09 & 21.36    & 13.10         & {{{13.17}}}         & {\bf{{23.82}}}         & 9.49      & 2.25$\times$ \\
&{\tt{europe\_osm}} & \text{\sffamily X}           & 50M & 108M & 1573.62 & 942.14 & 417.25 & 481.92   & 160.11        & {{207.46}}        & 136.98        & 136.98    & 3.52$\times$ \\
&{\tt{com-Friendster}} &  $\checkmark$          & 65M & 3.6B & 346.95 & (error) & (error) & 135.61   & 145.47        & 50.93         & 113.07        & 51.69     & 2.62$\times$ \\
&{\tt{Spielman\_k600}} &  $\checkmark$          & 72M & 216M & 520.89 & 40.46 & 231.24 & 181.34     & 129.43        & 130.83        & 14.41         &{\textit{\textbf{20.38}}}     & 8.90$\times$ \\
&{\tt{webbase-2001}} & $\checkmark$            & 118M & 1.0B & 291.98 & 72.81 & 74.70 & 29.23     & 25.02         & 12.44         & \bf{{49.29}}         & {\textit{\textbf{14.02}}}     & 2.08$\times$ \\
&{\tt{kmer\_V1r}} &  $\checkmark$               & 214M & 465M & 1493.65 & 59.98 & 244.60 & 317.91   & 361.76        & 81.67         & 166.08        & 43.19     & 7.36$\times$ \\
&{\tt{mawi}} &  $\checkmark$                    & 226M & 480M & 272.86 & 882.87 & 107.07 & 457.30   & 42.42         &{{46.43}}         & \bf{{57.39}}         & {\textit{\textbf{64.11}}}     & 7.13$\times$ \\
\midrule\midrule
\multicolumn{3}{c}{Spedups are computed over} &\multicolumn{2}{|c||}{Min}       & 0.10$\times$ & 0.07$\times$ & 0.22$\times$ &           1.0$\times$  &0.88$\times$&  1.39$\times$& 0.59$\times$ & \multicolumn{2}{c}{2.08$\times$} \\
\multicolumn{3}{c}{the graphs the corresponding} &\multicolumn{2}{|c||}{Max}       & 1.68$\times$ & 5.30$\times$ & 4.27$\times$ &           1.0$\times$  &10.78$\times$&  9.85$\times$& 12.58$\times$ & \multicolumn{2}{c}{8.90$\times$} \\
\multicolumn{3}{c}{kernel could process.} &\multicolumn{2}{|c||}{Geomean}           & 0.27$\times$& 0.77$\times$  & 0.73$\times$ &          1.0$\times$  &1.89$\times$  & 2.71$\times$ &  1.98$\times$&  \multicolumn{2}{c}{3.58$\times$} \\
\multicolumn{14}{c}{}
\end{tabular}}
\caption{\small
Performance (in milliseconds) of \acro against state-of-the-art BFS implementations, with \acro evaluated in four configurations: \acro (a) is the base variant with the proposed BVSS and optimal TC layout. \acro (ac) augments \acro (a) with the lazy vertex-update scheme. \acro (ab) augments \acro (a) with the reordering approach using window size $w = 2^{16}$. Finally, \acro (full) is the final algorithm, incorporating all the optimisations (a), (b), and (c) proposed throughout the paper.  All runtimes are in milliseconds (averaged over 64 runs), and the next column compares \acro against~\cite{berrybees}, the only other kernel using Tensor Cores and being the fastest in the literature.}
\label{tab:blest_vs_sota}
\end{table*}

\acro with lazy vertex updates is given in Algorithm~\ref{alg:bvss_kernel_lazy}. In the lazy marking phase~(Stage 1), there is exactly one \emph{potential} update per a pulling vertex, and this update is fully asynchronous. Furthermore, during the frontier-finalisation phase~(Stage 2), all the updates are fully coalesced: each thread processes a block of 32 vertices via 32-bit memory reads. Only the threads with at least one assigned vertex requiring a frontier push proceed further in the pipeline. Note that with \(\sigma = 8\), four slice sets reside in such a 32-bit block. When a thread finds a vertex requiring an update~($diff \neq 0$), it must iterate over all four slice sets (the four 8-bit chunks) and push the VSSs~[$start, end$) of any lazily marked vertices it encounters. The threads also set the vertices' levels to the current BFS level \(\ell\) at which the update takes place. Within each $\sigma$-bit mask, we identify the vertices of the next frontier by sparsely traversing the 8-bit slice chunk using the \textbf{\_ffs} instruction, which returns the position~(starting from 1) of the least significant set bit in the mask. Finally, to push the VSSs of the next level to \(\textbf{Q}_{\text{next}}\), a full atomic (\textbf{ATOMG}) is issued for space allocation in a warp-aggregated manner: only one thread per warp, having accumulated the total range of all threads, increments the queue size, and the resulting contiguous segment is then partitioned among the threads in that warp. This reduces the cost of the only remaining full atomic by a factor of 32, down to 1. 

Although lazy marking/update introduces an ${\Theta}$\((n)\) loop per BFS level, the fact that we opt for it only for social-like networks largely mitigates this concern. These networks have inherently low diameter, and therefore, the total number of these full passes is very small. For instance, the average number of BFS levels processed for GAP-twitter~\cite{suitesparse, gap} with $61.6$ million vertices is roughly 15.

During the frontier-finalisation phase (Stage~2), each thread, inspecting 32 consecutive vertices, may find that none requires an update, causing it to wait for the other warp threads. Fortunately, this scenario is rare in social networks, where frontiers are densely populated. Returning to GAP-twitter as an example, a randomly selected source visits 35{,}016{,}138 vertices in 15 levels, corresponding to an average of 2{,}334{,}409 frontier vertices per level. Given that there are 1{,}924{,}325 32-bit words in the frontier representation, a 32-bit word likely contains at least one set bit; words with no active vertices are extremely uncommon when the frontiers are crowded.

\subsection{Fusing Kernels}
\label{sec:kernel_fusing}
\acro employs a kernel-fusion optimisation to eliminate the need for host-side synchronisation between BFS levels. Traditionally, BFS on GPUs—whether in top-down or bottom-up form—relies on the host for level synchronisation, since CUDA before 9.0 lacked device-wide, grid-level synchronisation primitives. With the introduction of Cooperative Groups in CUDA 9~\cite{cooperative_groups}, device-wide synchronisation becomes possible within a single kernel launch. We leverage this capability to fuse all level kernels into a single persistent kernel, so that control never returns to the host between levels. This device-side synchronisation significantly accelerates BFS, especially on road networks that may require processing thousands of BFS levels.

\section{Experiments}
\label{sec:experiments}

\acro\footnote{\url{https://github.com/delbek/blest}} is implemented in C++/CUDA and compiled with \texttt{g++}~12.3.0 and \texttt{CUDA}~13.0. It supports all NVIDIA GPUs starting from compute capability~8.0 (Ampere) and onwards, i.e., all TC-equipped GPUs supporting the \textit{m8n8k128} instruction. The experiments are conducted on a system equipped with 2$\times$ Intel Xeon Gold 6152 CPUs, each having 22 cores (44 cores and 88 hardware threads with hyperthreading in total) and 1~TB RAM on the host side, and an NVIDIA H200 GPU with 141~GB of memory on the device side.

Table~\ref{tab:blest_vs_sota} shows the performance of \acro against the state-of-the-art BFS implementations, namely the CPU-based parallel BFS kernel of {\small GAP}~\cite{direction_optimized_bfs}, and GPU-based kernels, Gunrock~\cite{Wang:2017:GGG}, {\small GSWITCH}~\cite{gswitch}, and BerryBees~\cite{berrybees}. In the experiments, we used two benchmarks: the GAP Benchmark Suite~\cite{gap} and a custom suite containing all graphs from SuiteSparse~\cite{suitesparse} with $|\mathcal{V}| \ge 23M$ and $|\mathcal{E}| \le 2^{32} - 1$~(we removed {\tt{\small road\_usa}} and {\tt{\small sk-2005}} from the latter since the former includes them). All timings are averaged over 64 BFSs, executed from the same set of randomly selected roots.

Overall, {{\em{\ul{$\acro$ is faster than BerryBees $3.63\times$ on the GAP benchmark and  $3.58\times$ on the custom, large-graph benchmark}}}}. This being the most important result, the experiments reveal other insights and suggest opportunities for future work.

First, {{\em{\ul{every optimisation contributes to the final performance.}}}} The \emph{baseline}, \acro~(a), is faster than SoTA in 12/14 graphs, and on average $1.89\times$ faster. Although both the (ac) and (ab) variants, which are $2.71\times$ and $1.98\times$, respectively, are faster than SoTA, improve the base variant's performance, none can reach the performance of the (full) variant. Hence, each optimisation is useful. 

{{\em{\ul{\acro~(full) is faster than all the frameworks for all the benchmark graphs with average speedups of $13.25 \times$, $4.64 \times$, and $4.90 \times$ compared to GAP, Gunrock, and GSWITCH, respectively.}}}} However, \acro's static pipeline, which helps to obtain this superiority, and also, the best performance among its variants in general, could not choose the best one in 4/14 graphs.\footnote{If it could, the speedup would be $3.91\times$ instead of $3.58\times$.} These 4 cases are marked in bold and italic in the (full) column. Unlike the literature, e.g., {\small GSWITCH}, \acro has a single frontier representation which combines a bitwise array and a queue, and a single TC-based pull policy. Yet, the update mechanism can be lazy or not. Even when it is, the same policy is applied to each BFS level. That is, unlike SoTA, no on-the-fly analysis and dynamic decision making is performed. We leave this as a promising feature work. 

Among the 4/14 graphs on which the (full) variant fails to have the best performance, 3 suffer from reordering. A closer investigation reveals that when compared with the base variant (a), the (ab) variant worsens the performance, in fact, for 6/14 graphs, which are shown in bold in the (ab) column, including the three mentioned above. On the contrary, for the remaining eight graphs, reordering improves the performance with the largest benefit observed on {\tt{\small{Spielman\_k600}}}, where ordering yields an $8.98\times$ speedup. This duality prompted a more detailed analysis, from which we concluded that the {\em{\ul{high quality of the \emph{natural} orderings of these graphs masks the benefits of reordering}}}. To support this claim, we perform two experiments; the 6 worst-performing graphs are first randomly reordered, and second, the proposed reordering is used on them. The results of \acro (a) under this random ordering against \acro (ab)  are reported in Table~\ref{tab:random_vs_ordering}. 

\begin{table}[htbp]
\centering
\setlength{\tabcolsep}{2pt}
\scalebox{0.9}{
\begin{tabular}{lrr|lrr}
& \multicolumn{1}{l}{\textbf{Rnd. Order}} & \multicolumn{1}{l}{\textbf{\acro{}}}  & & \multicolumn{1}{l}{\textbf{Rnd. Order}} & \multicolumn{1}{l}{\textbf{\acro{}}} \\
\textbf{Graph} & \multicolumn{1}{l}{\textbf{\acro{} (a)}} & \multicolumn{1}{l|}{\textbf{(ab)}} & \textbf{Graph} & \multicolumn{1}{l}{\textbf{\acro{} (a)}} & \multicolumn{1}{l}{\textbf{(ab)}} \\
\midrule
{\tt{GAP-web}}               & 108.86 & 27.30 & {\tt{it-2004}}               &  76.16 & 23.82  \\
{\tt{webbase-2001}}          & 108.46 & 49.29 & {\tt{mawi}}                  &  57.37 & 57.39 \\
{\tt{uk-2005}}               &  64.46 & 21.42 & {\tt{nlpkkt240}}             &  64.26 & 17.41 \\
\multicolumn{6}{c}{}
\end{tabular}}
\caption{\small
Effect of replacing the natural ordering by a random reordering on graphs where \acro (ab) initially appears worse than \acro (a). Runtimes are in milliseconds.
}
\label{tab:random_vs_ordering}
\end{table}

Except for {\tt{\small{mawi}}}, \acro (a) with random ordering performs 
worse than \acro (ab); on average, it is $2.60\times$ slower, showing that the proposed ordering approach can be highly effective yet may be inferior to natural orderings. It is worth noting that 4 out of these 6 graphs are web graphs collected via web crawls; during a crawl, pages are typically discovered in a BFS-like manner and are labelled in approximately BFS order. As a result, {{\em{\ul{natural orderings may or may not be quite favourable while generating desirable sets in BVSS}}}}. A preliminary assessment of the matrix in its natural form within the pipeline would be beneficial, which we leave as future work.     

\begin{figure}[t]
  \centering
  \includegraphics[width=\linewidth]{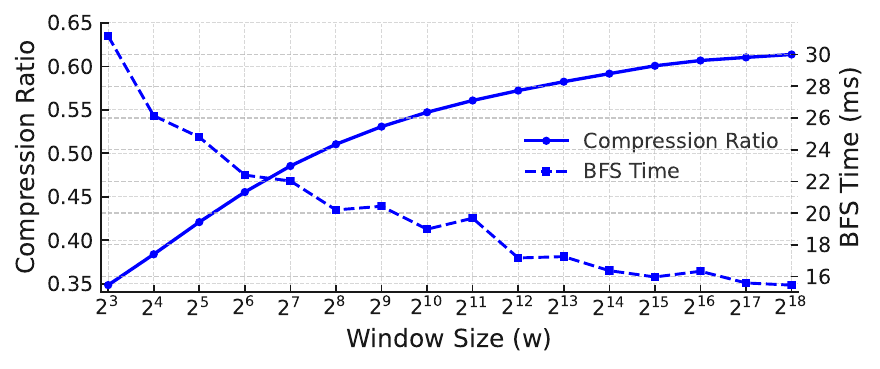}
  \caption{\small Effect of window size \(w\) on compression ratio and BFS runtime on GAP-web.}
  \label{fig:jaccard-window-compression-time}
\end{figure}

To empirically verify the soundness of the {\sc JaccardWithWindows} algorithm, we measure the effect of $w$ on the compression ratio and BFS time in Figure~\ref{fig:jaccard-window-compression-time} on {\tt{\small GAP-web}}. In the previous experiments, we have used a fixed window size \(w = 2^{16}\) for all the graphs. In Fig.~\ref{fig:jaccard-window-compression-time}, we study how the ordering quality evolves as \(w\) changes. To this end, we run the reordering algorithm on \textit{GAP-web} several times while increasing the window size from \(w = 2^{3}\) to \(w = 2^{18}\), and report the compression ratio and BFS runtime. As expected, {{\em{\ul{increasing w consistently improves the compression and almost always reduces the BFS execution time.}}}} The effective compression ratio increases from \(0.34\) at \(w = 2^{3}\) to \(0.61\) at \(w = 2^{18}\), resulting in an overall BFS speedup of \(2.01\times\). {{\em{\ul{The improvement, however, is clearly concave-down, indicating diminishing returns.}}}}

\begin{table*}[htbp]
\setlength{\tabcolsep}{2pt}
\centering
\scalebox{0.93}{
\begin{tabular}{c|l|rrrrrrr|rrr|r}
& \textbf{} &
\textbf{} &
\textbf{\#Virtual} &
\textbf{} &
\textbf{\#Unpadded} &
\textbf{\#Connectivity} &
\textbf{Update} &
\textbf{Comp.} &
\multicolumn{4}{c}{\textbf{Memory (GB)s}} \\
& \textbf{Graph} &
\textbf{\#Slice Sets} &
\textbf{Slice Sets} &
\textbf{\#Slices} &
\textbf{Slices} &
\textbf{Bits (unpadded)} &
\textbf{Divergence} &
\textbf{Ratio} &
\textbf{BVSS} & 
\textbf{Dynamic} & 
\textbf{Level} & 
\textbf{Total} \\ 
\midrule
\multirow{5}{*}{\rotatebox[origin=c]{90}{GAP}}
&{\tt{GAP-road}}              & 2,993,419 & 2,993,419 & 383,157,632 & 48,228,067 & 385,824,536 & 896 & 0.14 & 1.93 & 0.04 & 0.09 & 2.06 \\
&{\tt{GAP-twitter}}           & 7,697,302 & 14,596,630 & 1,291,994,772 & 1,281,499,157 & 10,251,993,256 & 2,547,849 & 0.14 & 6.54 & 0.14 & 0.25 & 6.93 \\
&{\tt{GAP-web}}               & 6,329,519 & 6,820,453 & 406,763,716 & 397,672,461 & 3,181,379,688 & 3,127,176 & 0.60& 2.08 & 0.07 & 0.21 & 2.36 \\
&{\tt{GAP-kron}}              & 16,777,216 & 43,995,265 & 4,246,770,008 & 4,221,563,999 & 33,772,511,992 & 5,387,729 & 0.12 & 21.47 & 0.40 & 0.54 & 22.41 \\
&{\tt{GAP-urand}}             & 16,777,216 & 41,041,325 & 4,291,771,932 & 4,266,603,115 & 34,132,824,920 & 4,455,452 & 0.12& 21.69 & 0.37 & 0.54 & 22.60 \\
\midrule
\multirow{10}{*}{\rotatebox[origin=c]{90}{$|\mathcal{V}| \ge 23M$, $|\mathcal{E}| \le 2^{32} - 1$}}
&{\tt{nlpkkt240}}             & 3,499,200 & 4,448,349 & 385,496,772 & 381,594,812 & 3,052,758,496 & 1,807,396 & 0.24 & 1.95 & 0.05 & 0.11 & 2.11 \\
&{\tt{uk-2005}}               & 4,932,491 & 5,081,525 & 213,416,292 & 206,111,967 & 1,648,895,736 & 2,707,017 & 0.55 & 1.10 & 0.06 & 0.16 & 1.32 \\
&{\tt{it-2004}}               & 5,161,450 & 5,501,639 & 248,213,140 & 240,644,435 & 1,925,155,480 & 2,585,730 & 0.58 & 1.28 & 0.06 & 0.16 & 1.50 \\
&{\tt{europe\_osm}}           & 6,364,003 & 6,364,003 & 814,592,384 & 101,677,915 & 813,423,320 & 982 & 0.13 & 4.12 & 0.07 & 0.20 & 4.39 \\
&{\tt{com-Friendster}}        & 8,201,046 & 32,010,653 & 3,518,615,984 & 3,506,329,253 & 28,050,634,024 & 1,556,318 & 0.12& 17.75 & 0.28 & 0.26 & 18.29 \\
&{\tt{Spielman\_k600}}        & 9,022,551 & 9,025,134 & 1,155,217,152 & 144,405,579 & 1,155,244,632 & 71,806 & 0.12& 5.84 & 0.10 & 0.29 & 6.23 \\
&{\tt{webbase-2001}}          & 14,767,770 & 14,036,757 & 333,225,540 & 311,783,859 & 2,494,270,872 & 6,948,498 & 0.39 & 1.78 & 0.15 & 0.48 & 2.41 \\
&{\tt{kmer\_V1r}}             & 26,750,628 & 26,750,628 & 3,424,080,384 & 440,760,418 & 3,526,083,344 & 1,547,870 & 0.13& 17.33 & 0.29 & 0.86 & 18.48 \\
&{\tt{mawi}}                  & 28,274,524 & 30,088,197 & 357,289,860 & 277,409,693 & 2,219,277,544 & 276,100 & 0.21& 2.01 & 0.33 & 0.91 & 3.25 \\
\multicolumn{13}{c}{}\\
\end{tabular}}

\caption{\small The structural statistics and memory footprint of BVSS across evaluated graphs, reporting the number of virtual vs. unpadded slice sets and slices, the total unpadded connectivity bits, the update-divergence metric, the effective compression ratio, and the GPU memory breakdown (GB) for BVSS~(static), dynamic arrays~(frontier/visited), and the level array. The last column shows the total memory in GBs.}
\label{tab:bvss_statistics}
\end{table*}

Returning to Table~\ref{tab:blest_vs_sota}, the lazy update scheme (c) is substantially effective on almost all graphs for which the library decides to apply it~(on top of (ab)). This indeed significantly accelerates \acro. For the two graphs where the lazy vertex update \emph{degrades} performance, namely {\tt{\small mawi}} and {\tt{\small Spielman\_k600}}, we uncover another important detail. These graphs exhibit relatively small update divergence values of 276,100 and 71,806, respectively\footnote{As a reference, the update divergence of GAP-twitter is 2,547,849.}. We used a fixed threshold of 25,000 for \acro to switch to the lazy vertex update scheme. Hence, {{\em{\ul{a better tuning mechanism to apply the lazy vertex scheme may improve performance.}}} Moreover, the BFS on \textit{Spielman\_k600} takes 600 levels to complete, each including a $\Theta(n)$ pass, as required by the mechanics of the lazy update scheme, which is specifically designed for social-like networks with few levels. As mentioned above, we plan to develop a sophisticated decision pipeline that accurately determines which policy each graph should be submitted to, in order to prevent such anomalies.

For completeness, Table~\ref{tab:bvss_statistics} reports the BVSS data-structure footprint; we emphasise that it is compact enough to fit within the memory budgets of a wide range of GPUs—from mainstream devices to high-end accelerators—thereby enabling scalable deployment without imposing prohibitive storage overhead.

\section{Related Work}
\label{sec:related_work}

The idea of using linear-algebraic primitives in graph algorithms transformed graph analytics~\cite{combblas, graphblas, graphblast}. This became prominent with the availability of MMA units in modern accelerators. BerryBees~\cite{berrybees} is an early effort to realise BFS using TCs via a bitmap-based data structure. In contrast to \acro, it uses both SpMV and SpMSpV as a primitive and switches from the latter to the former whenever the frontier is sufficiently dense\footnote{This switching mechanism is different from the direction-optimising scheme proposed by Beamer et al.~\cite{direction_optimized_bfs}. It does not switch between push and pull. Instead, it attempts to improve the work efficiency by skipping pulls for vertices whose incoming neighbours contain no frontier vertices, i.e., it realises an early exit in SpMSpV.}. Both variants of their implementation, however, suffer from major bottlenecks as explained in this work.
BerryBees' SpMV applies a \emph{frontier-oblivious slice-set distribution}, which in turn severely exacerbates load imbalance across the warps. When the frontier becomes sparse, BerryBees switches to SpMSpV. However, this control logic does not align with their unit of work, i.e., \emph{slice set}. This switching should be based on the number of active sets, rather than the number of active vertices.

BerryBees tries to solve the load-balancing problem in the SpMSpV by traversing the frontier and redistributing the slice sets based on the number of slices. This per-level, frontier-dependent redistribution limits the benefits of SpMSpV and highlights the importance of the proposed BVSS data structure, which achieves near-perfect load balance \emph{by construction}, without requiring the algorithmic layer to be aware of or to react to such dynamic load variations.

Gunrock~\cite{Wang:2017:GGG} is a GPU library designed to balance performance, scalability, and programmability by expressing the graph algorithms through a data-centric “frontier” abstraction and a
bulk-synchronous loop. It applies GPU-specific optimisations such as load-balanced workload mapping, idempotence to reduce atomics under concurrent discoveries, and push vs. pull direction switching. Hence, it specialises the execution flow with respect to the graph at hand. The literature, and also this work, follows the message that there is no single “best” GPU BFS frontier expansion policy that does not vary w.r.t. the graph structure. GSWITCH~\cite{gswitch} dynamically reconfigures (with low overhead) the execution by selecting among optimisation “patterns,” most critically direction switching (push vs. pull), plus complementary choices such as load-balancing strategy and active-set representation. It uses runtime/workload features to decide when to switch and which kernels to run, searching a large BFS variant space each iteration. \acro (currently) has a preliminary, static execution specialisation, and even with this, it can be the fastest thanks to the BVSS and its optimal layout on Tensor Cores. We leave the design of a better static/dynamic kernel-selection pipeline as future work. 

\section{Conclusion and Future Work}
\label{sec:conclusion}

This paper proposes \acro, a highly efficient data structure and algorithm for computing BFS using Tensor Cores on modern GPUs. We introduced various optimisations, each targeting a bottleneck of an existing TC-based BFS framework, and demonstrated that these contributions yield an average speedup of \(3.58\times\) over the state-of-the-art. The results, without a doubt, position \acro as a milestone in efficiently exploiting Tensor Cores for graph algorithms.

As future work, we will extend \acro to graph analytics tasks based on multi-source BFSs, e.g., closeness centrality and diameter computation, both in single- and multi-GPU configurations. We also plan to port \acro to AMD GPUs, also equipped with MMA units that can support \acro's execution pipeline. As discussed before, for certain graphs, we inadvertently select the lazy-update scheme, which indicates that a more sophisticated decision pipeline is needed to determine when the lazy vertex update should be enabled. In addition, we will design techniques that dynamically manipulate the execution flow based on the current BFS state.

\section{Acknowledgements}
\label{sec:acknowledgements}
The numerical calculations reported in this paper were partially performed at TUBITAK ULAKBIM, High Performance and Grid Computing Center (TRUBA resources).

\balance
\bibliographystyle{plain}
\bibliography{main}

\end{document}